\documentclass{WileyMSP-template}

\begin{document}

\pagestyle{fancy}
\rhead{\includegraphics[width=2.5cm]{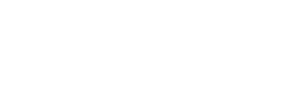}}

\title{Area Selective Chemical Vapor Deposition of Gold by Electron Beam Seeding}

\maketitle


\author{Aleksei Tsarapkin*}
\author{Krzysztof Maćkosz}
\author{Chinmai Sai Jureddy}
\author{Ivo Utke}
\author{Katja Höflich*}


\dedication{}

\begin{affiliations}
A. Tsarapkin, Dr. K. Höflich\\
Ferdinand-Braun-Institut (FBH), Leibniz-Institut für Höchstfrequenztechnik, Gustav-Kirchhoff-Str. 4, D-12489 Berlin, Germany\\
Email Addresses: aleksei.tsarapkin@fbh-berlin.de, katja.hoeflich@fbh-berlin.de

Dr. K. Maćkosz, C. S. Jureddy, Dr. I. Utke\\
Empa - Swiss Federal Laboratories for Materials Science and Technology, Feuerwerkerstrasse 39, CH-3602 Thun, Switzerland

\end{affiliations}



\begin{abstract}

Chemical vapor deposition (CVD) is an established method for producing high-purity thin films, but it typically necessitates the pre- and post-processing \textcolor{black}{using} a mask to produce structures. 
This paper presents a novel maskless patterning technique that enables area selective CVD of gold.
A focused electron beam is used to decompose the metal-organic precursor Au(acac)Me$_2$ locally, thereby creating an autocatalytically active seed layer for subsequent CVD with the same precursor.
The procedure \textcolor{black}{could} be included in the same CVD \textcolor{black}{process} without the need for clean room lithographic processing.
Moreover, it operates at low temperatures of 80~\textdegree C, over 200~K lower than standard CVD temperatures for this precursor, reducing thermal load on the specimen.
Given that electron beam seeding operates on any even moderately conductive surface, the process does not constrain device design.
This is demonstrated by the example of vertical nanostructures with high aspect ratios of around 40:1 and more. 
Written using a focused electron beam and the same precursor, these nanopillars exhibit catalytically active nuclei on their surface.
Furthermore, \textcolor{black}{by using the onset of the autocatalytic CVD growth}, for the first time the local temperature increase caused by the writing of nanostructures with an electron beam \textcolor{black}{could be precisely determined}.

\end{abstract}


\section{Introduction}

Gold is a key metal in microelectronics due to its low resistivity, resistance to electromigration and chemical inertness.  
These properties make it an excellent choice for stable and reliable electrical contacts even under challenging environmental conditions~\cite{Baum1994}. 
Chemical vapor deposition (CVD) is a tech\-nologically significant option for fabricating gold contacts or other metallic functional structures.
CVD utilises the thermal decomposition of \textcolor{black}{adsorbed} precursor molecules \textcolor{black}{delivered} in a gas\-eous state~\cite{Hampdensmith1994}. 
Molecules reaching the surface are thermally decom\-posed, leaving the central metal atom on the surface. 
If the decomposition is favored kinetically or thermodynamically on certain surfaces compared to others, the ensuing variation in reaction rates can be exploited for area selective deposition~\cite{Hampdensmith1994}.
This has been shown for various metals, e.g. Au on W~\cite{Colgate1990}, Au on Cr~\cite{Banaszakholl1993}, or Cu on Pt~\cite{Lecohier1992}. 
All of the aforementioned approaches require sophisticated lithography methods to define the growth surfaces, which can restrict the selection of substrate materials, ultimately constraining the degrees of freedom in device design~\cite{Colgate1990, Gladfelter1993}. 
Consequently, an area selective deposition technique that operates without the need for lithography, and is independent of the substrate type, would be highly advantageous.

One option is to provide a localized energy input for CVD, which is frequently accomplished through laser irradiation~\cite{Baum1985, Irvine2008}. 
However, the spatial resolution achievable through this method is typically restricted by the wavelength of the laser light employed.
Although there has been recent evidence that this issue can be addressed in thermally decoupled 3D gold structures~\cite{Lasseter2023}, the range of achievable structure geometries remains restricted.  
This is also applicable to local energy input through resistive heating~\cite{Porrati2023}. 
An alternative approach uses beams of charged particles, often electrons~\cite{Kunz1987-1, Kunz1987-2}.

Instead of supplying additional energy during the CVD process, electrons can also be used to activate a specific surface~\cite{Walz2010, Muthukumar2012} or to deposit a seed layer for preferential decomposition of precursor molecules either by CVD~\cite{Kunz1987-2, Kunz1988} or by atomic layer deposition (ALD)~\cite{Mackus2010, Mameli2019}. 
Seeding using a focused electron beam possesses significant potential, as direct electron beam writing is capable of operating on any surface topography \textcolor{black}{and} even \textcolor{black}{on} weakly conductive \textcolor{black}{materials}~\cite{Utke2008}. 
Additionally, it provides the possibility for unparalleled spatial resolution~\cite{vanDorp2008}.
If the seeds previously deposited exhibit autocatalytic activity~\cite{Kunz1987-2, Kunz1988, Vollnhals2014, Martinovic2023}, the temperature necessary for decomposition decreases considerably.
\textcolor{black}{Here, the term autocatalytic CVD is used to highlight that seeds of the same metal enhance the reaction rate for thermal decomposition which} efficiently limits the CVD metal growth to a specific area, while also reducing the thermal load on the substrate. 
However, there is currently no proof of electron beam seeding resulting in autocatalytic growth for gold.

\begin{figure*}[ht]
\centering
\includegraphics[width=16cm]{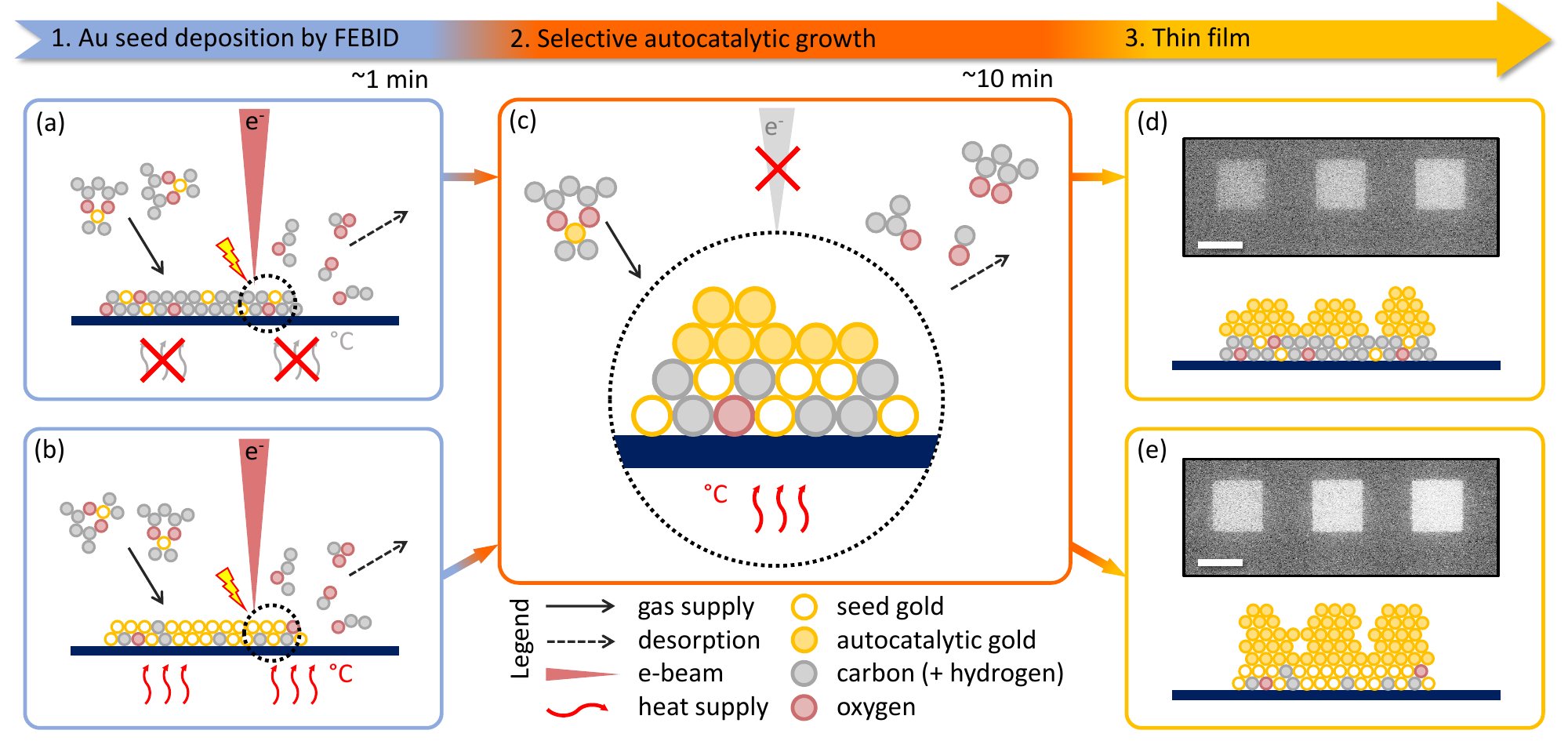}
\caption{Schematic representation of the area selective autocatalytic film growth. A seed layer is deposited using a focused electron beam to locally decompose the precursor on a substrate (a) at room temperature (30~\textdegree C) or (b) at elevated substrate temperatures. (c) After seeding the precursor is decomposed at the deposited Au nuclei without further electron-beam assistance. SE micrographs show the resulting gold films for (d) room-temperature seeding and (e) seeding at elevated substrate temperatures, both proving the area selectivity of the process. Scale bars represent 2~\textmu m.}
\label{fig:fig1}
\end{figure*}

In this study, area selective CVD of gold by electron beam seeding is demonstrated for the first time.
The process employs a focused electron beam for the non-thermal electron-triggered decomposition of Au(acac)Me$_2$ which results in autocatalytically active gold nuclei. 
Subsequently, the same precursor is thermally decomposed at the gold seeds. 
This autocatalytic CVD of gold is observed at relatively low substrate temperatures of about 80~\textdegree C and requires only high vacuum conditions.
Electron beam seeding is highly versatile, as demonstrated by the deposition of high aspect ratio pillars using the same precursor. 
The tempera\-ture-depen\-dent autocatalytic CVD in this case provides the first quantitative measurement of the temperature change induced by the electron beam during the direct writing of \textcolor{black}{high aspect-ratio} structures.

\section{Results and Discussion}

The principle of the proposed process is depicted in Figure~\ref{fig:fig1}. 
A seed layer is deposited by the impact of a focused electron beam on a sample surface at room or elevated temperature. 
The electron beam is then switched off while the molecular gas flow is maintained for a period of time. 
Subsequent investigations show autocatalytic growth of pure gold at both seeding temperatures.

\subsection{Electron Beam Seeding}

\begin{figure*}[ht] 
\centering
\includegraphics[width=16cm]{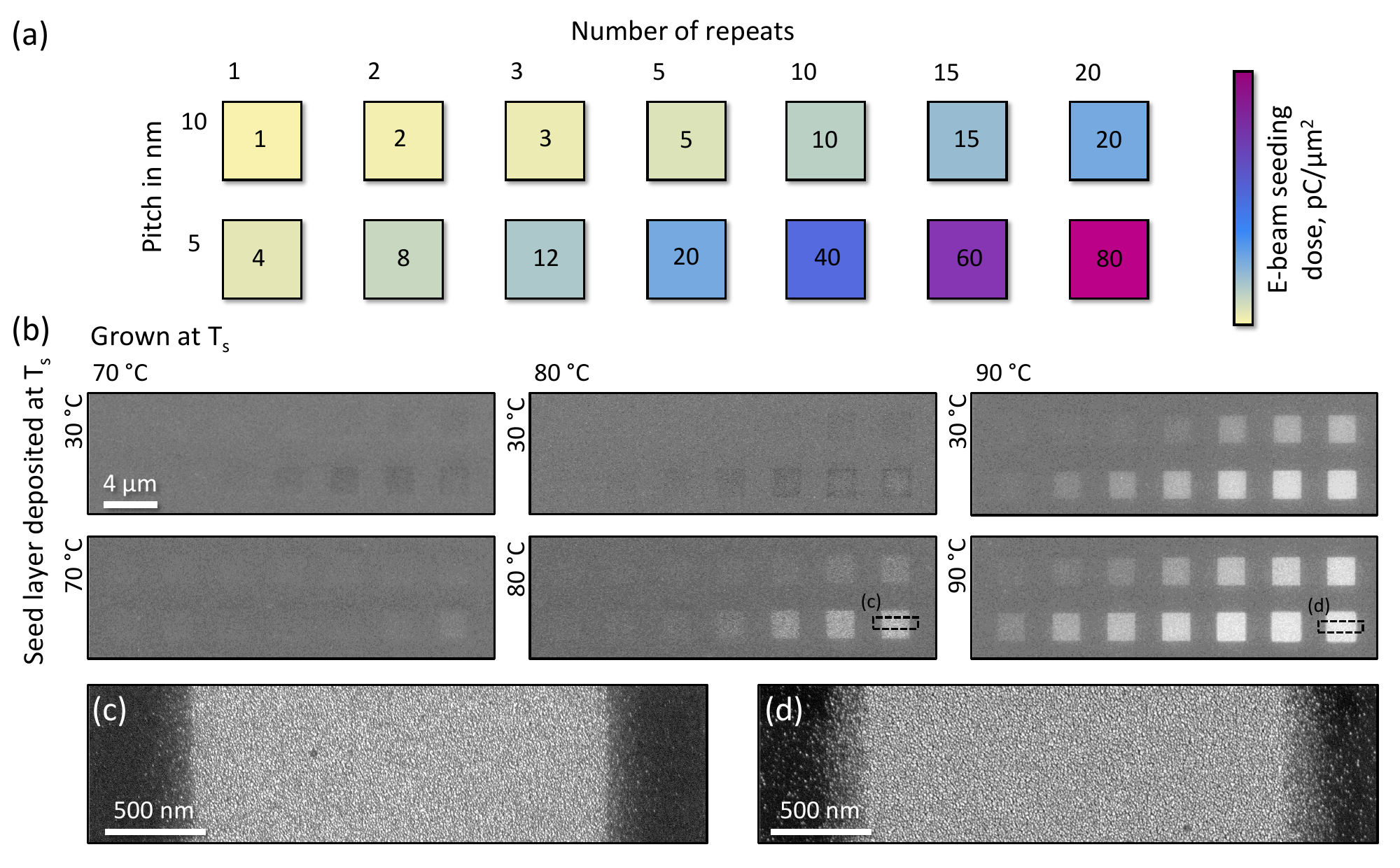}
\caption{
(a)~Experimental seed layer deposition array. The non-linear colour scale represents the \textcolor{black}{electron seeding dose applied} for a single square. (b)~SE micrographs of the gold films grown on the seed arrays, pre-deposited on a stage at room temperature ($\approx$ 30~\textdegree C) (top row) and on a hot stage (bottom row). The gold films were grown autocatalytically for 30 min at elevated temperature, varied from 70 to 90~\textdegree C. \textcolor{black}{High resolution SE close-up of selected regions of two squares, grown and seeded (c) at 80~\textdegree C and (d) at 90~\textdegree C .}}
\label{fig:fig2}
\end{figure*}

All experiments were carried out in a conventional dual beam instrument fitted with a gas injection system (GIS) to introduce precursor molecules locally and a sample heating stage, as detailed in the Supporting Information (SI).
The autocatalytically enhanced CVD process was found to work reliably on all used bulk substrates. 
Nevertheless, it is crucial to clean the sample surface prior to the process. 
Any organic or inorganic residues on the surface hinder the reproducibility of the process by partially impeding autocatalysis, or by functioning as preferred precursor chemisorption sites and thus "contamination seeds."

The precursor compound used, Au(acac)Me$_2$; dimethyl\-(acetyl\-acetonate)\-gold(III), is primarily known for its use in CVD where it provides pure gold deposition in plasma assisted processing at low temperatures~\cite{Feurer1987}.  
The thermal CVD process was carried out at a substrate temperature of 300~\textdegree C. 
It produced gold films of medium quality with 85~at.\% gold content. 
However, the corresponding fluori\-nated dike\-tonates, namely tfac and hfac (tetra- and hexa\-fluoro\-ace\-to\-nate), resulted in pure gold deposition~\cite{Larson1986}.
Au(acac)Me$_2$ is also commonly utilized in focused electron beam induced deposition (FEBID), a direct writing technique where the precursor molecules physisorbed to the surface are dissociated using a focused electron beam~\cite{Utke2008}.

\textcolor{black}{E}lectron-induced dissociation \textcolor{black}{is mainly triggered by secondary electrons with an energy distribution between 0 and 50 eV. Therefore, direct electron beam writing} is not bond selective \textcolor{black}{and} the use of metal-organic precursor compounds often results in a carbon-rich de\-po\-sit~\cite{Utke2008, Botman2009}.
Efforts have been made to deposit materials with inorganic ligands, which have shown promise especially for magnetic materials~\cite{Barth2020}. 
For gold, the inorganic precursor, Au(CO)Cl, containing both chlorine and carbonyl ligands, allows high-purity deposition using focused electron beam writing. 
However, the chlorine content restricts spatial selectivity due to etching effects~\cite{Mulders2012}.
Similar to PF$_3$AuCl, which also provides high-purity gold deposition~\cite{Utke2000}, \textcolor{black}{Au(CO)Cl} is not stable under vacuum conditions, severely limiting its practical usability~\cite{Marashdeh2017}.
It would be of significant interest to explore gold carbonyls in this context. 
Although they can be synthesized, gold carbonyls exist only in specific laboratory environments and are therefore not suitable for typical FEBID or CVD experiments~\cite{McIntosh2002, Jiang2005}.
For this reason, precursor compounds with organic ligands were used in most of the reported FEBID experiments~\cite{Utke2022}.

The key advantage of the direct electron beam writing process is that it provides extreme shape flexibility in three dimensions~\cite{Winkler2019}.
To efficiently utilize these distinctive 3D capabilities, several techniques have been employed to attain pure deposits.
Here, we will particularly focus on the influence of heat, either as beam-induced heating, especially for large beam currents and thin 3D geometries, or applied externally by sample heating at the various stages of processing.
Heat input is viewed as a means of enhancing deposit purity as it promotes the desorption rates of cleaved ligands and contributes to ligand dissociation by thermal assistance~\cite{Utke2022}.
However, the acac ligand has a low desorption rate and a tendency to polymerize under electron beam impact. 
This limits the achievable gold contents to approximately 30~at.\% when using thermal energy input~\cite{Sqalli2002, Mulders2011}.
To achieve pure 3D gold structures, additional oxidizing agents must be used in combination with elaborate shape optimization for effective purification~\cite{Hoeflich2011, Winkler2017, Kuhness2021}.
The key question is whether this presents a fundamental constraint, or whether thermal assistance at low electron fluxes, as in the case of silver deposition~\cite{Hoeflich2017, Martinovic2022}, can result in pure yet still selective metal deposition. 
In this scenario, the clean surface of the gold seed could even be autocatalytically active.

To investigate this hypothesis, the following experimental matrix was performed. 
Under a continuous gas flow, an array of 2x2~\textmu m$^2$ squares was patterned with varying electron beam doses using a beam energy of 5~keV and a beam current of 100~pA. 
Figure~\ref{fig:fig2}(a) shows part of the array, the complete array is available in the SI.
\textcolor{black}{The applied electron dose in pC/\textmu m$^2$ is provided within each square.} 
The corresponding pixel spacing (pitch) and number of repetitions (repeats) are also indicated alongside. 
The utilized pitches of 10 and 5 nm correspond to doses of 6.25 and 25 electrons per nm$^2$ and \textcolor{black}{frame} repeat, respectively.
Starting from a single repeat which is supposed to deposit only isolated gold atoms and small clusters up to 100 repeats were conducted, as the latter would lead to a thin FEBID deposit.
\textcolor{black}{In addition, the effective local dose changes when the pitch is varied, as the dwell time is kept constant. With the same total dose, seed layers may exhibit different purities as co-dissociation of ligands or a too slow ligand desorption rate could play a role \cite{SanzHernandez2017}}.
To investigate the effect of ligand desorption, the seed deposition was carried out at room temperature (30~\textdegree C) and at elevated stage temperatures.
Similar to early reports on Fe(CO)$_5$ by Kunz~\cite{Kunz1987-2}, we observed electron-enhanced CVD \textcolor{black}{i.e. an accelerated growth rate} through continued electron irradiation under low \textcolor{black}{electron} fluxes and elevated stage temperatures (see Figure~SI.4).
Therefore special care was taken to avoid unintended beam exposure, and no imaging was conducted before or after the seed array deposition.
As the GIS system has an automatically controlled valve, gas flow occurred only during seeding for the array deposition time of 1~min 18~s.
\textcolor{black}{Different} control experiments for electron beam induced surface activation and \textcolor{black}{for seeding using other precursors were performed} with the same array deposited at the different substrate temperatures \textcolor{black}{ (see Figures~SI.5-.7)}.

\subsection{Autocatalytic Growth of Gold}

In order to achieve autocatalytic activity of the seed layer, two conditions must be met.
Firstly, the previously deposited metallic nuclei must have a clean surface to participate in the reaction. 

Additionally, all final reaction products that result from ligand cleavage must be volatile enough to desorb intactly~\cite{Utke2022}. 
\textcolor{black}{Secondly, the growth kinetics at the nucleus must promote chemisorption of the precursor molecules through the removal of its ligands}~\cite{Kunz1988}.

This condition applies to some \textcolor{black}{metal} carbonyls, e.g. area selective CVD by electron beam seeding has been de\-monstrated for the deposition of transition metals using precursors like Fe(CO)$_5$~\cite{Kunz1987-2, Martinovic2023}, \newline Co(CO)$_3$NO~\cite{Vollnhals2014}, and Cr(CO)$_6$~\cite{Kunz1988}.
In some cases high vacuum (HV) conditions were sufficient to observe autocatalysis, e.g. for Fe(CO)$_5$~\cite{Hochleitner2008}, but in others clean metal surfaces were additionally required excluding electron beam writing for seeding, e.g. for Co$_2$(CO)$_8$~\cite{Cordoba2012}.
Under ultra high vacuum condi\-tions, NH$_3$ dosing was found to in\-hibit auto\-cata\-lytic growth, which could sub\-sequent\-ly be reactivated by electron beam seeding, pro\-vi\-ding an additional control option in area selective CVD utilising Fe(CO)$_5$~\cite{Martinovic2023}.
\textcolor{black}{As the carbonyl ligand is} thermodynamically stable and \textcolor{black}{neutral}, \textcolor{black}{carbonyls are} the preferable choice for autocatalytically enhanced growth processes. 
\textcolor{black}{However, as discussed above gold carbonyl precursors are not vacuum stable.} 
\textcolor{black}{Gold precursors typically contain non-neutral} organic ligands such as the frequently used $\beta$-diketonates (here acac) \textcolor{black}{which have less} favorable desorption properties~\textcolor{black}{\cite{Madey1986}}. 


The autocatalytic growth experiments were performed in the same vacuum chamber. 
After depositing a square array for one \textcolor{black}{given} temperature, the electron beam was switched off and the GIS nozzle was opened again for 30~min to deliver precursor molecules.

Scanning electron (SE) micrographs for investigation were taken only after cooling the stage to room temperature and recovering the background pressure to the base pressure of about $6\cdot10^{-7}$ mbar. 
The experimental matrix continued with the next array, covering temperatures ranging from 30~\textdegree C up to 90~\textdegree C at intervals of 10~K.

The growth results are shown in Figure~\ref{fig:fig2}(b) where carbon appears as a dark contrast and gold as a bright contrast, which is a typical characteristic of SE imaging.
At 60~\textdegree C, no observable difference was noted when compared to the seed arrays deposited initially. 
However, at 70~\textdegree C, initial signs of CVD growth became evident. 
The contrast shifted from dark (carbon-rich) to bright (gold-rich), becoming more pronounced as the growth temperature increased.
Squares with a greater number of repeats in the electron beam seeding are brighter, corresponding to more pronounced gold growth.
This can be attributed to an increased density of gold nuclei available with deposition time.
The optimal balance between growth rate and selectivity is achieved at a temperature of 80~\textdegree C\textcolor{black}{, as shown in a close-up Fig.~\ref{fig:fig2}(c)}. 
Raising the temperature further enhances the growth rate, but decreases selectivity due to the initiation of growth in the halo regions around the actual patterned squares. 
This phenomenon is demonstrated in Figure~\ref{fig:fig2}\textcolor{black}{(d)}, a high-resolution SE micrograph of the square seeded at 90~\textdegree C with 20 repetitions at 5~nm pitch.
Although there are no primary electrons in the halo, there are still secondary and back\-scattered electrons present with a significantly lower flux than in the deposition region~\cite{Hoeflich2017}.
Consequently, numerous repetitions in seeding at 80~\textdegree C can also diminish the precision of square edges.

Interestingly, electron beam seeding also works at room temperature.
However, as anticipated \textcolor{black}{from~\cite{Mulders2011}}, the higher desorption rate of ligands during seeding at elevated stage temperature suppresses carbon co-deposition, resulting in higher densities of catalytically active seeds. 
This is illustrated in Figures~\ref{fig:fig1}(a) and (b).
Therefore, the films grown from seed layers deposited at room temperature show carbon \textcolor{black}{as low} contrast \textcolor{black}{areas}, while at the elevated stage temperature the seed density is already high enough to form a quasi-continuous gold film, cf. Figure~\ref{fig:fig2}(b) middle row. 

For the tested pitch values and numbers of repeats, the growth of the autocatalytic film at a certain temperature is only influenced by the total electron dose.
When comparing SE contrast and topography, it is evident that the seeds, which were deposited with different pitches and repeats but the same total dose, display the same thickness and SE contrast (grain size) after growth\textcolor{black}{, cf. Fig.~\ref{fig:fig2}(b) for 20~pC/\textmu m$^2$ and SI for more examples}. 
The optimum dose is reached at about 140~pC/\textmu m$^2$ or 875 electrons per nm$^2$.

Additionally, the process exhibits exceptional selectivity as gold growth is restricted to the presence of gold nuclei under these temperatures.
It is worth noting, however, that a substantial decrease in CVD growth rate was observed over time, resulting in saturation for attainable thicknesses at approximately 6.5~nm \textcolor{black}{(cf. SI for more details)}. 
This could be due to surface poisoning \textcolor{black}{\cite{Cordoba2012}}, i.e. carbon coverage of the active gold nuclei due to residual hydrocarbons in the vacuum chamber and/or low desorption rates of the ligand fragments.
In this instance, an improvement in the background vacuum would be promising.
Another potential reason is the nuclei changing in size during autocatalytic growth.
Size effects can have a significant impact on the catalytic activity of metallic nanoparticles, especially in the case of gold~\cite{Haruta1997}.
It is worth mentioning that the observed decreasing CVD growth rate is not an obstacle to technological implementation, as seeding and growth can be easily iterated several times (as shown in Figure~SI.4).

\begin{figure}[tb] 
\centering
\includegraphics[width=10cm]{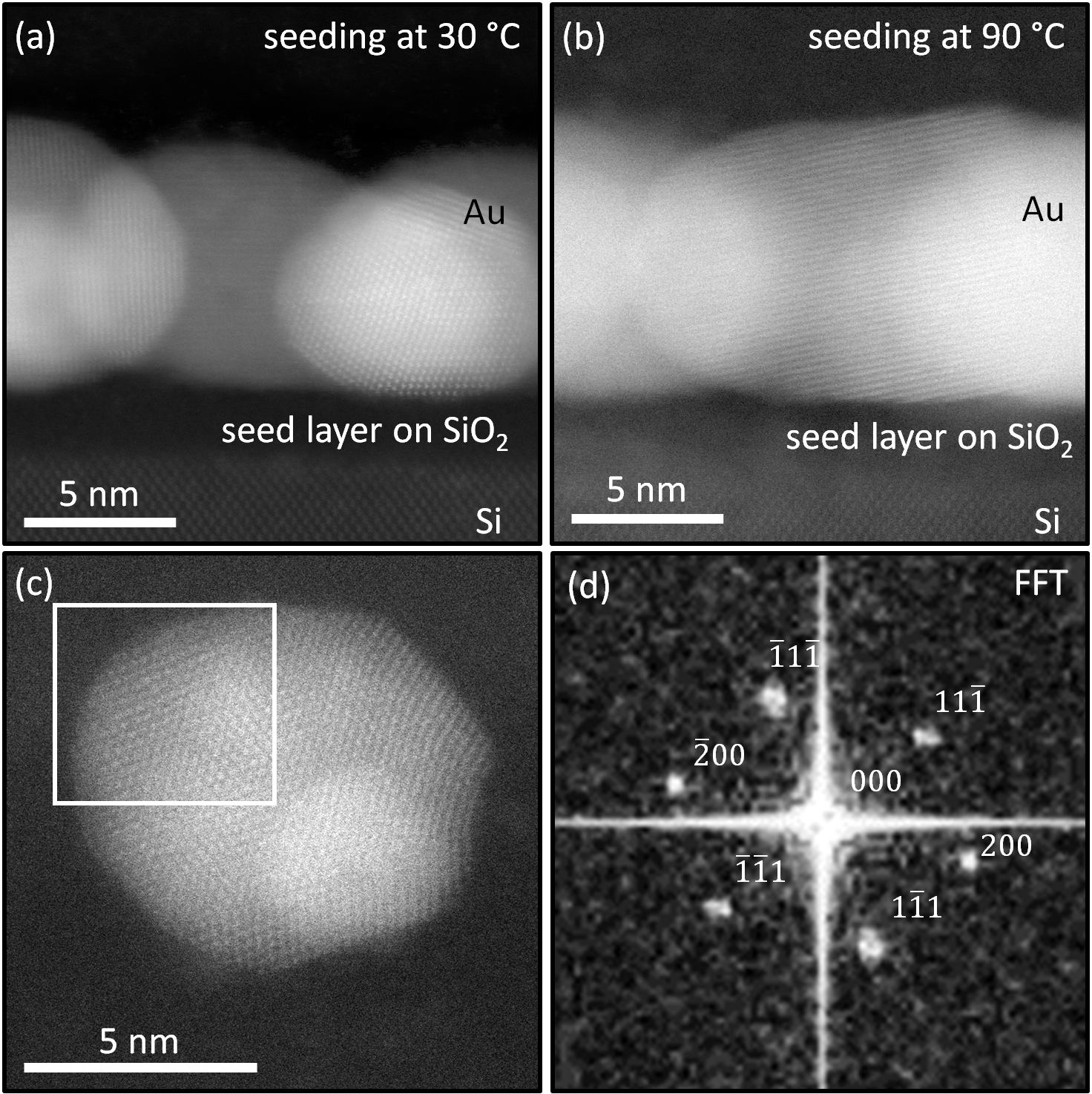}
\caption{STEM HAADF images of gold films grown on a native oxide silicon substrate at 90~\textdegree C. (a) Seed layer deposition on a cold stage at 30~\textdegree C, (b) Seed layer deposition at 90~\textdegree C, and (c) a single gold particle from halo region with (d) FFT pattern of the selected region of interest confirming a single-crystalline gold structure with a lattice constant of 4.073~\AA.}
\label{fig:fig3}
\end{figure}

The quality of the gold films obtained was later assessed by scanning transmission electron microscopy (STEM).
Figure~\ref{fig:fig3} depicts cross-sectional images of films seeded at (a) 30 and (b) 90~\textdegree C, both grown at 90~\textdegree C.
The high-angle annular dark-field (HAADF) images prove the presence of pure crystalline gold in both instances, recognizable as the periodic atomic order in bright contrast regions. 
The close-up displayed in Figure~\ref{fig:fig3}(c) of a distinct grain from the halo region permits the evaluation of the crystal structure through the acquisition of a fast-Fourier transform (FFT) image that can be indexed based on the lattice parameters of single-crystalline gold.

\subsection{Beam-induced Heating and Autocatalysis}

One of the main benefits of using a tightly focused electron beam to decompose molecules from the gas phase is its extreme 3D flexibility, resolution and shape fidelity~\cite{Winkler2019}.
Generally, FEBID is considered to be stimulated by secondary electrons which dissociate the physisorbed precursor molecules.
Electron-induced chemistry involves various reaction pathways, which depend on the local availability of electrons, molecules, and co-reactants~\cite{Barth2020, Utke2022}.
Beam-induced heating is also a significant factor.
As the growth surface locally heats up, the dynamics of the supplied precursor molecules and fragments will shift, e.g. towards faster diffusion and higher desorption rates. 
Estimates show that even slight changes in surface temperature, around 10~K, can significantly impact the vertical growth rate during 3D FEBID~\cite{Randolph2005, Mutunga2019, Skoric2020, Fowlkes2023}.
However, there is no experimental data available regarding the actual local temperature changes \textcolor{black}{at the growing nanopillar apexes where the e-beam impacts}.
Here, we exploit the onset of autocatalytic CVD during the growth of one-dimensional vertical nanostructures (pillars) under steady-state growth conditions as a means of in situ temperature measurement.

\begin{figure}[tb]
\centering
\includegraphics[width=10cm]{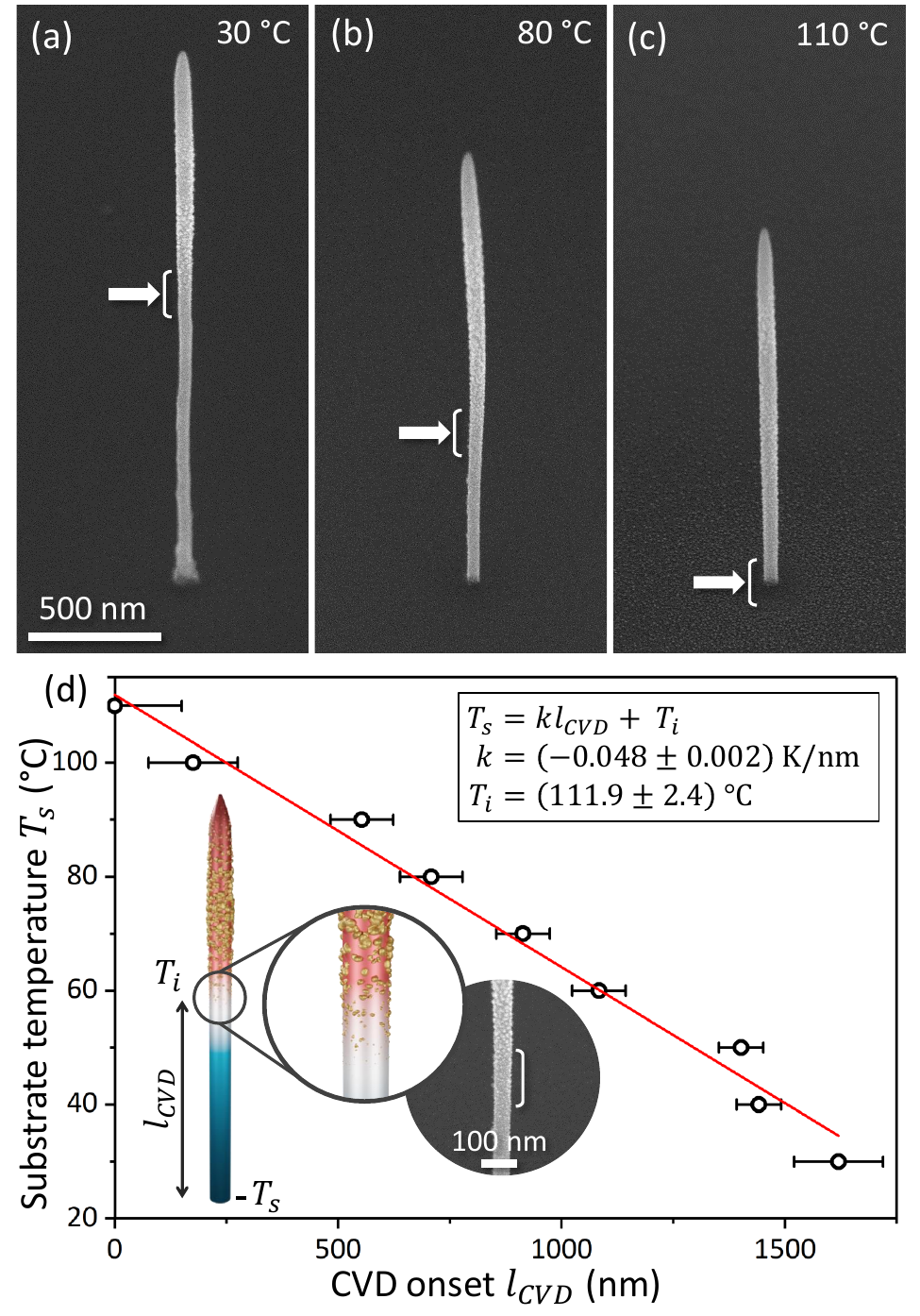}
\caption{
FEBID grown pillars deposited on the substrate at elevated temperatures, (a) 30~\textdegree C, (b) 70~\textdegree C, and (c) 110~\textdegree C. (d) - CVD onset height vs. sample temperature. The colour scale in the inserted sketch of the pillar represents the linear temperature distribution during FEBID.}
\label{fig:fig4}
\end{figure} 

Figure~\ref{fig:fig4}(a)-(c) shows example images of deposited nano\-pillars using the precursor Au(acac)Me$_2$ with diameters of about 50~nm and aspect ratios of 40:1 and larger. 
The primary \textcolor{black}{electron} energy used was 15~keV at a beam current of 100~pA, which aligns with typical standards in 3D FEBID printing \textcolor{black}{\cite{Skoric2020, Fowlkes2016, Keller2018}}.
Stationary growth conditions were achieved using a dwell time of 26.2~ms~\cite{Randolph2005, Fowlkes2016}.
Each pillar's overall growth duration was consistently maintained at 2~min 30~s deposition time.

Figure~\ref{fig:fig4} (a) presents the nanopillar obtained at room temperature.
\textcolor{black}{It shows a base cylinder with smooth surface and} bright \textcolor{black}{contrast} gold particles in the upper part.
The presence of gold particles confirms the existence of autocatalytically active gold nuclei on the surface of the pillar and the attainment of the temperature threshold T$_i$ for visible particle growth.
\textcolor{black}{No CVD growth was observed on the remaining nanopillar surface below and the surface retained its original smooth appearance.} 
By adjusting the substrate temperature, the region of particle growth onset can be moved along the pillar, as demonstrated by the white arrows in Figures~\ref{fig:fig4}(a)-(c).
The corresponding substrate temperatures are subsequently plotted as a function of the vertical distance $l_{\text{CVD}}$ from the substrate surface to the onset region, cf. Figure~\ref{fig:fig4}(d).
The resulting decrease in $l_{\text{CVD}}$ with the increase in substrate temperature is linear.

This linear relationship is in line with the fact that there is a negligible loss of thermal energy across the surface of the pillar (cf. SI for estimates on radiative and convective heat losses).
Therefore, all the thermal energy introduced by the stationary electron beam is conducted through the pillar into the substrate, which can be considered as a constant temperature heat sink.
The heating induced by the beam during growth results in a linear temperature gradient within the pillar~\cite{Utke2008}. 
Therefore, the vertical position of the CVD onset region is dependent on the stage temperature in a linear manner. 
The overall height of the pillars reduces for higher stage temperatures, as a result of the increased desorption rate of precursor molecules. 

The intersection point of the linear fit with the temperature axis yields a measured temperature of T$_i \approx $ 112~\textdegree C, which indicates the point at which particle growth becomes visible.
The thermal decomposition of the precursor conforms to Arrhenius' law, with the autocatalysis rate exponentially dependent on temperature $\exp\,(-E_a / k_B T\big)$, where the activation energy $E_a$ acts as the threshold value and $k_B$ is the Boltzmann constant.
The overall growth duration was relatively short, requiring a significant growth rate to observe particle growth.
These findings align with our observations of the electron beam-seeded autocatalytic growth at temperatures of 100-110~\textdegree C, which was highly efficient but showed decreasing selectivity.
For growth \textcolor{black}{of high aspect-ratio structures}, reduced selectivity is not an issue since only the thermally decoupled areas reach high enough temperatures. 

The proportionality constant $k \approx 0.05$~K/nm indicates the temperature variation per unit length along the pillar.
It offers a quantitative measure of the thermal energy introduced by the electron beam, providing valuable insights for modelling 3D electron beam printing and the possibility of quantitatively determining the thermal conductivity of the deposited material.
For this purpose, the measurements are supported by Monte-Carlo modelling of the electron trajectories obtained for this geometry. 
A mass density range of $\rho = 4 - 6.5$~g$/$cm$^3$ is assumed, corresponding to the typically observed Au contents in deposits from this precursor, cf. SI for more details on the Monte-Carlo modelling.
The obtained range for the average inelastic energy loss $\Delta E_{\text{kin}} = 0.75 - 0.80~$keV  per incident electron inside the nanostructure can then be used to determine the thermal conductivity $\kappa$ of the deposit material:

\begin{equation}
    \kappa = \frac{I}{e}\frac{\Delta E_{\text{kin}}}{A}\frac{1}{k}  = 0.38 - 0.41 ~\frac{\text{W}}{\text{K$\cdot$ m}} \, ,
\end{equation}
with the electron beam current $I$, elementary charge $e$ and the cross-sectional area $A$ of the nanopillar.
This can be considered an upper limit since some inelastic losses are due to the production of secondary electrons (and X-rays), while the rest results in Joule heating (e.g. $\Delta E_{\text{Joule}} \approx 0.86 ~\Delta E_{\text{kin}}$ from an assessment of this geometry~\cite{Mutunga2019}).

The value $\kappa \approx$ 0.4~W/(K$\cdot$m) obtained for thermal conductivity
has an uncertainty margin of approximately 24~\% and falls within the commonly observed range for various carbon modifications, particularly those ranging from amorphous to diamond-like carbon~\cite{Robertson2002}.
By repeating this type of experiment for different acceleration voltages, beam currents and possibly other precursors, the thermal conductivity of the FEBID material for the different deposition conditions can be determined quantitatively.

\section{Conclusions \textcolor{black}{and Outlook}}

In this work, area selective CVD of gold by electron beam seeding is demonstrated for the first time.
Gold nuclei, acting as a seed layer, are deposited through a maskless procedure by means of a focused electron beam, followed by gentle heating of the substrate, which triggers autocatalytic gold growth only in the seeded regions.
The use of a focused electron beam ensures \textcolor{black}{high} spatial resolution, unhampered by mask usage typical in electron beam lithography and rendering complex multi-stage lithographic processing unnecessary.
Seeding and subsequent growth both use the same precursor compound, Au(acac)Me$_2$, which is commercially available, long-term stable and non-toxic.
The technique has been demonstrated to operate at low temperatures of approximately 80~\textdegree C, more than 200~K below the typical substrate temperature used in the established CVD process for this precursor.
Moreover, it demands solely a high vacuum, making technological implementation remarkably straightforward \textcolor{black}{for academic research prototyping}. 
Any commercially available scanning electron microscope can be used for this method if it is equipped with a gas injection system and a heating stage.
In the future, this process could be seamlessly integrated into conventional CVD reactors by equipping them with an electron gun featuring programmable position control.

Unlike many other area selective CVD processes, direct electron beam writing for seeding is not affected by the surface material or topography.
To demonstrate this capability, the same precursor was used to fabricate vertical nanostructures with high aspect ratios using the focused electron beam.
In addition to achieving autocatalytically active seeds \textcolor{black}{at high aspect ratio structures}, the first quantitative measurement of the temperature change during direct electron beam writing of \textcolor{black}{such} structures was performed. 
This quantitative determination of the up to now inaccessible local temperature evolution is a major breakthrough as it enables improved modelling for accurate shape 3D nanoprinting through FEBID within the additive manufacturing field at a sub-100~nm scale~\cite{Mutunga2019, Skoric2020, Fowlkes2023, Fowlkes2016, Keller2018}.

To summarise, this work presents a new, technologically powerful and easy to implement approach to area selective gold deposition.
The low temperature necessary makes it highly suitable for processing innovative flexible devices that rely on heat-sensitive polymers, including flexible electronics~\cite{Gates2009} and flexible solar cells~\cite{Pagliaro2008}.
Another exciting area of application is utilizing the maskless direct write seeding on 3D architectures, which enables the fabrication of nanostructures or localized electric contacts on intricate 3D devices.

\newpage

\section{Experimental Section}

Silicon wafers with native oxide and different doping levels were employed as the substrates and chipped into 1~cm x 1~cm pieces. 
As the doping level decides the conductivity and therefore the secondary electron yields, the minimum required electron doses for successful seeding varied. 
The silicon chips with a native oxide layer were vertically sonicated in N-methyl-2-pyrrolidone for 20~min, followed by a rinse with isopropyl alcohol.
The chips were then soaked in a 3:1 Piranha solution for at least 1~h before being rinsed with deionized water and dried using nitrogen gas.
This process effectively removes any organic material from the surface of the sample.

All testing was carried out using a Helios 5 UX dual beam microscope.
The gas injection system (GIS) with crystals of Au(acac)Me$_2$ (CAS: 14951-50-9) was heated to 30~\textdegree C.
To enable autocatalytic growth of gold films, gold nuclei were seeded with an electron beam at 5 kV acceleration voltage and a beam current of 100 pA. 
The base chamber pressure prior to opening the GIS valve was typically $6\cdot10^{-7}$ mbar, with the GIS valve open during autocatalysis $7.5\cdot10^{-7}$ mbar at the start of the experiments, dropping to approximately $7\cdot10^{-7}$ at the end. 
A Kleindiek vacuum compatible Micro Heating and Cooling Stage (MHCS) was used together with a temperature control system to heat the sample up to 110~\textdegree C.
Pillars for measuring the temperature induced by an electron beam were grown using a focused electron beam of 15~kV and 100~pA. 
Deposition for all pillars was achieved by repeating a single spot 5725 times using the maximum possible dwell time of 26.2~ms, resulting in a total deposition time of about 2~min 30~s for a single pillar.
The experimental matrices are fully described in the SI.

Lamellae for transmission electron microscopy (TEM) were prepared according to the standard lift-out protocol using the dual beam FIB-SEM system Tescan Lyra3.
The TEM images were captured using a probe-corrected ThermoFisher Scientific Titan Themis 200 G3.

\medskip
Supporting Information is available from the Wiley Online Library or 
from the author.

\medskip
\textbf{Acknowledgements} \par 
This work was funded by the German Research Association under grant agreement no. HO 5461/3-1 and by the Swiss National Fund by the COST-SNF project IZCOZ0\_205450.
We want to furthermore acknowledge support by the EU COST action CA 19140 'FIT4NANO' (www-fit4nano.eu). The authors are particularly grateful to Caroline Hain for her support with the English language\textcolor{black}{, and to Ibukun Olaniyan for the support with AFM measurements}.

\medskip

%


\newpage

\begin{figure}
\textbf{Table of Contents}\\
\medskip
  \includegraphics[width=11cm]{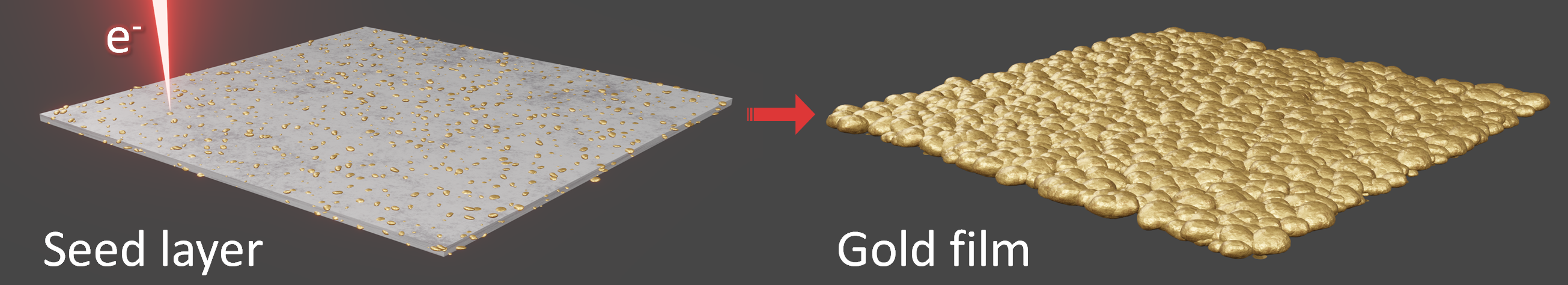}
  \medskip
  \caption*{In this work we demonstrate an area selective chemical vapor deposition of gold. A metal-organic precursor Au(acac)Me$_2$ is used to create an autocatalytically active seed layer by electron beam \textcolor{black}{deposition} for further film growth at sub-100~\textdegree C temperatures. In addition, this process \textcolor{black}{provides} a novel method for determining local temperature increases during electron beam writing.}
\end{figure}

\newpage
\section*{Supporting Information}
\setcounter{section}{0}
\renewcommand\thefigure{SI.\arabic{figure}}  
\section{Methods and Full Seed Array}

Silicon wafers with native oxide and different doping levels were employed as the substrates and chipped into 1~cm x 1~cm pieces. 
As the doping level decides the conductivity and therefore the secondary electron yields, the minimum required electron doses for successful seeding varied. 
The silicon chips with a native oxide layer were vertically sonicated in N-methyl-2-pyrrolidone for 20~min, followed by a rinse with isopropyl alcohol.
The chips were then soaked in a 3:1 Piranha solution for at least 1~h before being rinsed with deionized water and dried using nitrogen gas.
This process effectively removes any organic material from the surface of the sample.

All testing was carried out using a Helios 5 UX dual beam microscope.
The gas injection system (GIS) with crystals of Au(acac)Me$_2$ (CAS: 14951-50-9) was heated to 30~\textdegree C.
To enable autocatalytic growth of gold films, gold nuclei were seeded with an electron beam at 5 kV acceleration voltage and a beam current of 100 pA. 
The base chamber pressure prior to opening the GIS valve was typically $6\cdot10^{-7}$ mbar, with the GIS valve open during autocatalysis $7.5\cdot10^{-7}$ mbar at the start of the experiments, dropping to approximately $7\cdot10^{-7}$ at the end of an experimental day. 
A Kleindiek vacuum compatible Micro Heating and Cooling Stage (MHCS) was used together with a temperature control system to heat the sample up to 125~\textdegree C.

\begin{figure*}[ht] 
\centering
\includegraphics[width=16cm]{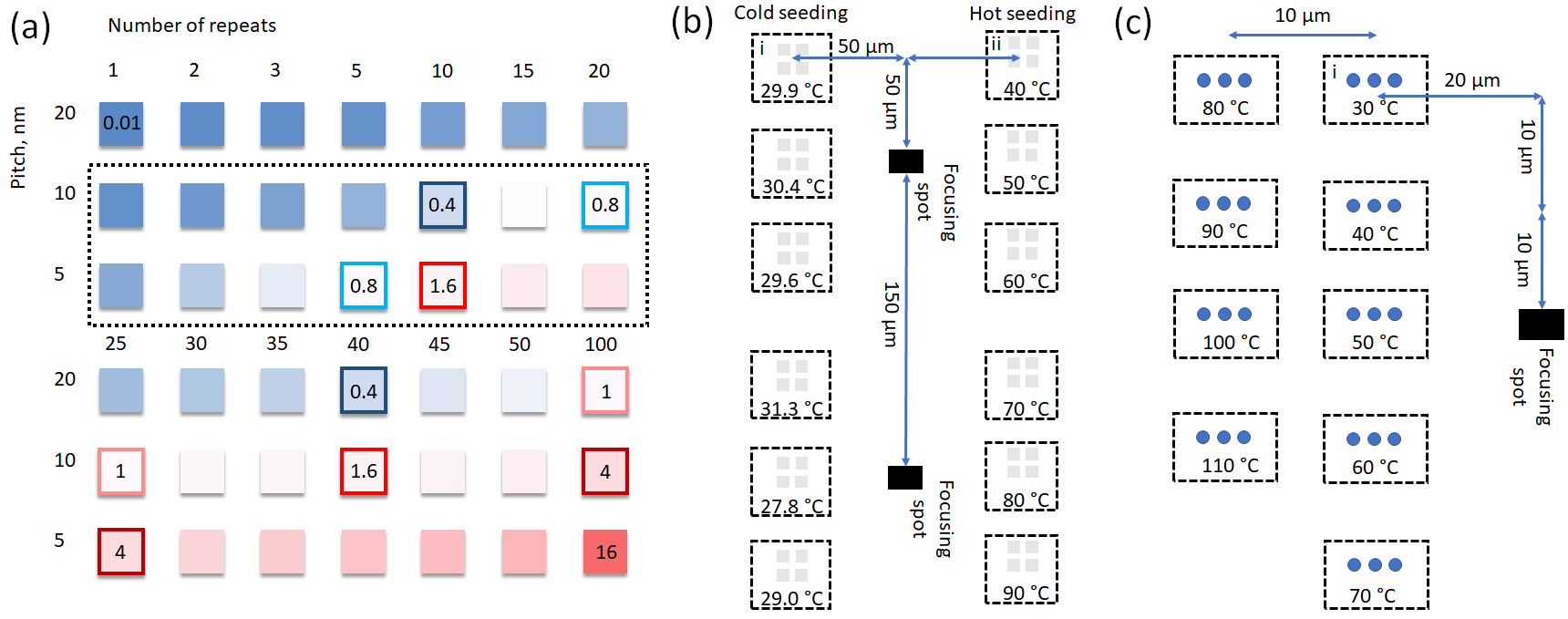}
\caption{(a) Schematic representation of the full seed array, the dashed line indicates the squares selected for the main manuscript. Each square has a deposition time varying between 10 ms and 16 s, with a total deposition time of 1 min 18 s for the full array. Squares with equal deposition times are highlighted with a colored border. An experimental matrix is provided for (b) the autocatalytic growth of gold films and (c) the deposition of vertical pillars to measure the temperature induced by the electron beam (T$_i$).}
\label{SIfig:fullmatrix}
\end{figure*}

The complete pattern array for all experiments is shown in Figure~\ref{SIfig:fullmatrix}(a). 
It comprises 2~\textmu m x 2~\textmu m squares with variable pixel-to-pixel pitches (5, 10 and 20~nm) of the electron beam, separated by 2~\textmu m, and different deposition times (achieved by scan repeats from 1 to 100). 
A non-linear color scale indicates the deposition time of a particular square, ranging from 10~ms to 16~s.
The whole seed array takes 1 min 18 s of deposition time to complete. 
The assortment of squares within the full seed array enables the gold film thickness to be optimised and to understand how autocatalytic growth depends on the electron beam patterning.
Squares featuring the same deposition time (and dose) are highlighted with a coloured frame.

The experimental layout of all autocatalytically grown films is shown schematically in Figure~\ref{SIfig:fullmatrix}(b). 
The complete experimental procedure for seeding and autocatalytic growth at various temperatures is detailed below:
\begin{enumerate}
    \item Focus the beam onto a designated area with a small feature on it. It is crucial to refocus the beam before each seed deposition owing to material expansion caused by heat.
    \item Move the stage to the first patterning area, which should be at least 50~\textmu m away from the focusing area (position i in Figure~\ref{SIfig:fullmatrix}(b)), without conducting any observation with the SEM.
    \item Deposit the full seed array at room temperature, approximately 30~\textdegree C for all experiments.
    \item Increase the stage temperature to 40~\textdegree C.
    \item Allow autocatalytic growth of the gold films on the seeded areas with an open GIS and maintain a stable stage temperature for 30 min.
    \item Returning the stage to the focusing area, refocusing the beam, and capturing a SE image of the resulting films.
    \item Moving to the second patterning area, which is situated 100~\textmu m from the previously grown films (position ii in Figure~\ref{SIfig:fullmatrix}(b)).
    \item Deposition of the full seed array onto the hot stage (40~\textdegree C).
    \item Autocatalytic growth of the gold films on the seeded areas with open GIS and stable stage temperature for a duration of 30 min.
    \item Capturing SE images of the resulting films.
    \item Returning to the focusing area and replicating the procedure for elevated temperatures ranging from 50 to 90~\textdegree C at intervals of 10~K
\end{enumerate}

Pillars for measuring the temperature induced by an electron beam were grown using a focused electron beam of 15~kV and 100~pA. 
Deposition for all pillars was achieved by repeating a single spot 5725 times the maximum possible dwell time of 26.2~ms, resulting in a total deposition time of about 2~min 30~s for a single pillar.
The layout for experimental deposition of all pillars under different stage temperatures is illustrated in Figure~\ref{SIfig:fullmatrix}(c).
The entire experimental protocol for pillars growth is described below:
\begin{enumerate}
    \item Focus the beam on a designated area with a small feature. Refocus the beam before each seed deposition due to thermal material expansion.
    \item Without observation with the SEM, move the stage to the first patterning area, which is at least 20~\textmu m away from the focusing area. Room temperature deposition is done at position i in Figure~\ref{SIfig:fullmatrix}(c).
    \item Deposit three pillars with the same parameters.
    \item Close and retract the GIS, tilt the stage to 45\textdegree.
    \item Take high-resolution SEM images of all pillars.
    \item Tilt and move back to the focusing area. 
    \item Repeat the same protocol for higher temperatures ranging from 40 to 110~\textdegree C with in steps of 10~K. 
    \item Ensure that all depositions are done in close proximity to the focusing spot to guarantee similar beam waist for each iteration. 
\end{enumerate}

\section{SE Micrographs of Autocatalytically Grown Films on the Full Seed Arrays}

\begin{figure*}[ht]
\centering
\includegraphics[width=17.4cm]{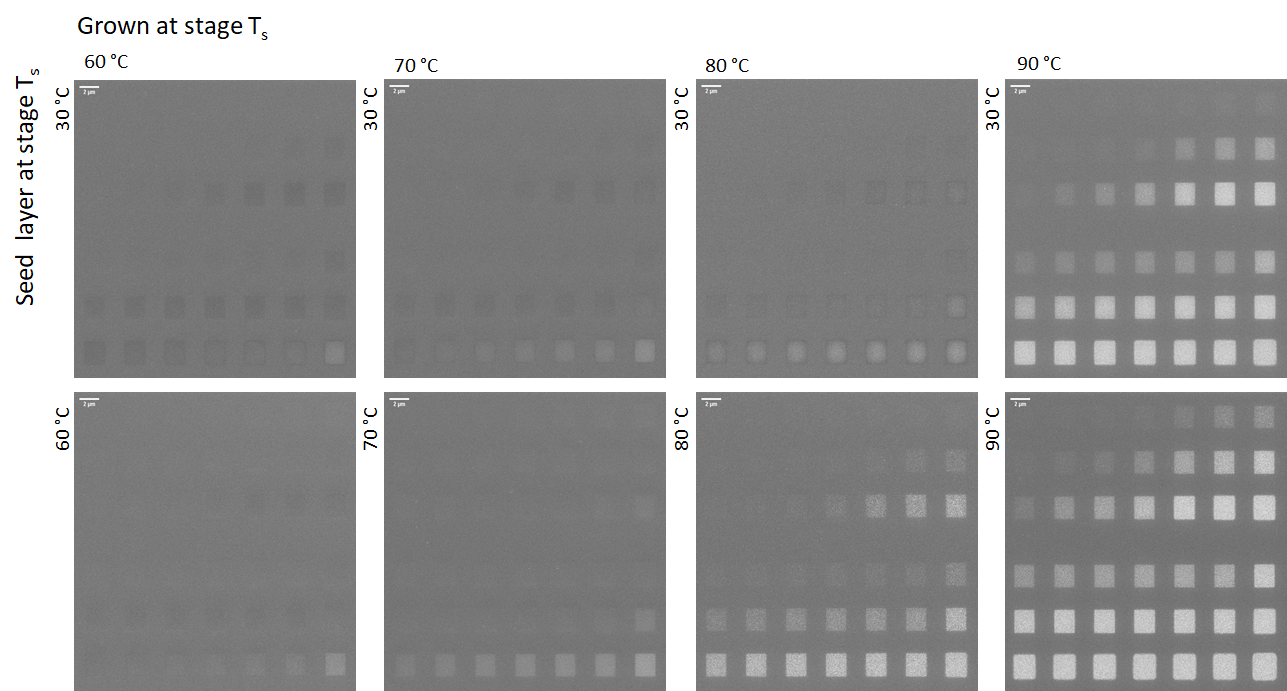}
\caption{SE micrographs of gold films grown autocatalytically by CVD for 30 min on the respective seed arrays. The vertically arranged temperature values indicate the substrate temperature (T$_s$) during the seed layer arrays deposition, whereas horizontally arranged temperatures specify T$_s$ for the subsequent autocatalytic Au CVD growth. The scale bars indicate 2~\textmu m.}
\label{SIfig:full1st}
\end{figure*}

SE micrographs of gold films, grown on the full seed arrays at various temperatures, are displayed in Figure~\ref{SIfig:full1st}. 
The gold films in the top row were grown on seed arrays deposited at a room temperature of 30~\textdegree C, while the seed arrays for the gold films in the bottom row were deposited at higher temperatures ranging from 60 to 90~\textdegree C. 
Autocatalytic growth was continued for 30 min for each film.

\begin{figure*}[ht]
\centering
\includegraphics[width=13cm]{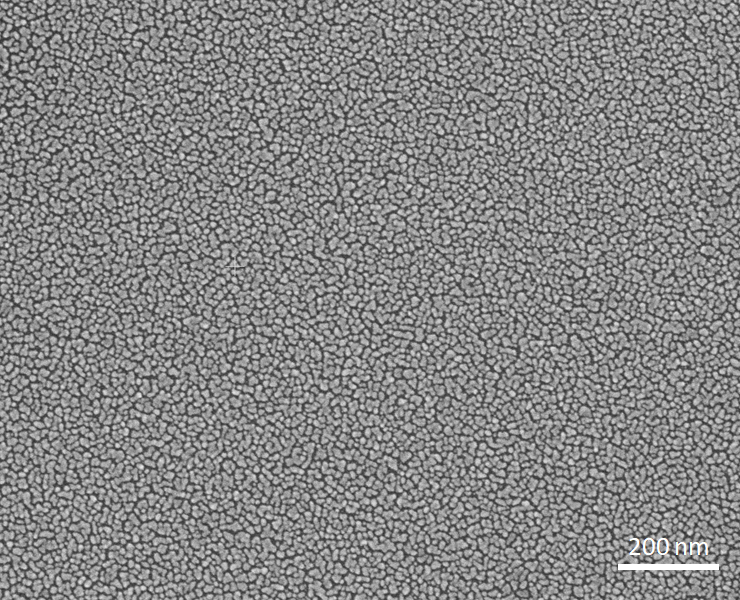}
\caption{\textcolor{black}{HR SE micrograph of gold film seeded at 90~\textdegree C substrate temperature and grown autocatalytically by CVD for 30 min at 90~\textdegree C.}}
\label{SIfig:grains}
\end{figure*}

\newpage
\textcolor{black}{High-resolution SE image analysis of a gold film seeded and grown at 90~\textdegree C (Fig.~\ref{SIfig:grains}) shows a lateral average grain size of 18~nm. 
Results are obtained with particle analysis plugin in open-source ImageJ software.
}

\newpage
\section{Time Evolution of Autocatalytic Growth on the Full Seed Arrays}

\begin{figure*}[ht]
\centering
\includegraphics[width=17.4cm]{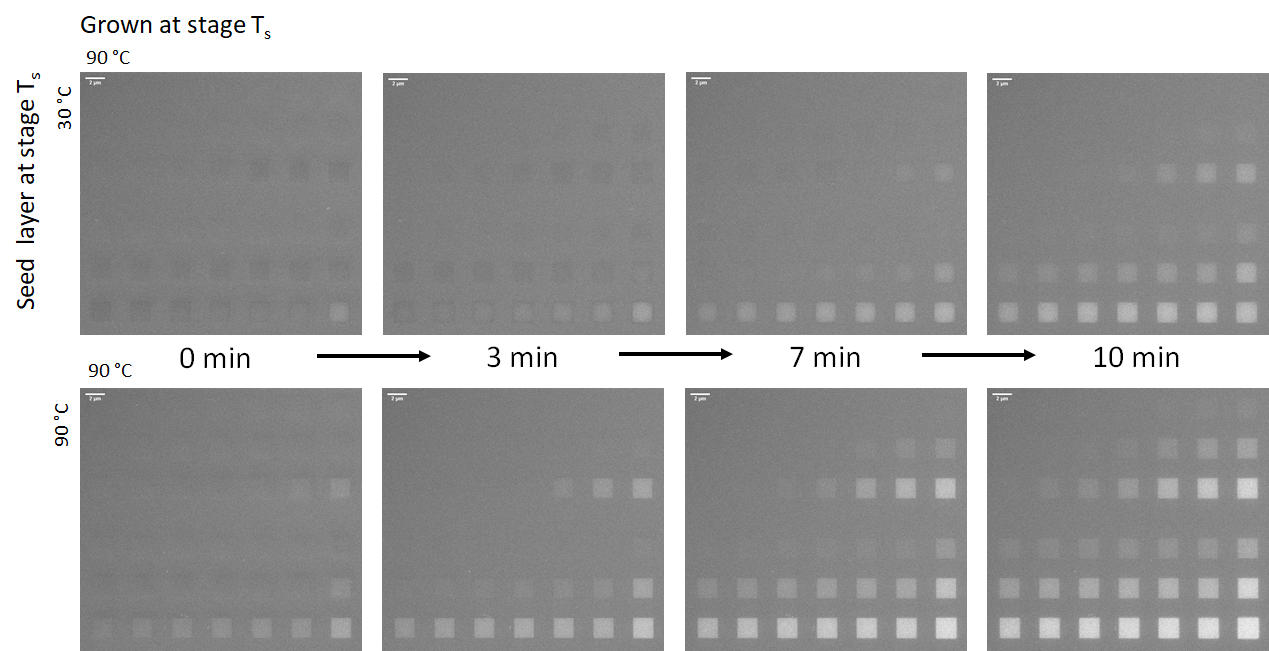}
\caption{SE micrographs of the gold films showing the time evolution of the autocatalytic growth by CVD on the respective seed arrays. The substrate temperature T$_s$ during seed array deposition for the upper row is 30~\textdegree C, while it is 90~\textdegree C for the lower row. Autocatalytic CVD growth for both arrays was done at 90~\textdegree C. The images are captured at 0, 3, 7, and 10 mins following the autocatalytic CVD growth.}
\label{SIfig:full2nd}
\end{figure*}

The autocatalytic growth of gold films on two seed arrays for additional intermediate electron supply is shown in Figure~\ref{SIfig:full2nd}, indicating the time evolution.
\textcolor{black}{Both autocatalytic growth of gold on the seeds and electron beam irradiation processes are simultaneously involved in this process (similar to \cite{Kunz1987-2})}.
The initial SEM image at 0 minutes in the top left corner shows the as-deposited seed array at room temperature (30~\textdegree C), taken immediately after heating to 90~\textdegree C. 
With the GIS open and stage heated for 10~min, several images of this array were captured every 60 seconds. 
The images were taken with a field of view of 40~\textmu m x 26.6~\textmu m and a dwell time 15~\textmu s. 
Thus, 100~pA electron beam current resulted in a dose of 0.022~pC/\textmu m$^2$ or 0.14 electrons per nm$^2$ per single image.
Figure~\ref{SIfig:full2nd} displays SEM images captured at 3, 7, and 10 mins to visualize the evolution of the autocatalytic gold CVD film growth.
The bottom row presents the as-deposited seed array, which was grown on the sample at an elevated temperature of 90~\textdegree C, with the same sequence of images taken at different times.
These images show that SE imaging acts as iterative seeding to enhance the CVD growth rate significantly compared to the pure autocatalytic growth demonstrated above.
Note that when imaging, seeds are also deposited between squares, resulting in a decrease in area-selectivity.
By implementing an iterative seeding approach, electrons will only be supplied in the selected areas.

\newpage
\section{Control \textcolor{black}{Experiments}}

\begin{figure}[ht]
\centering
\includegraphics[width=17.4cm]{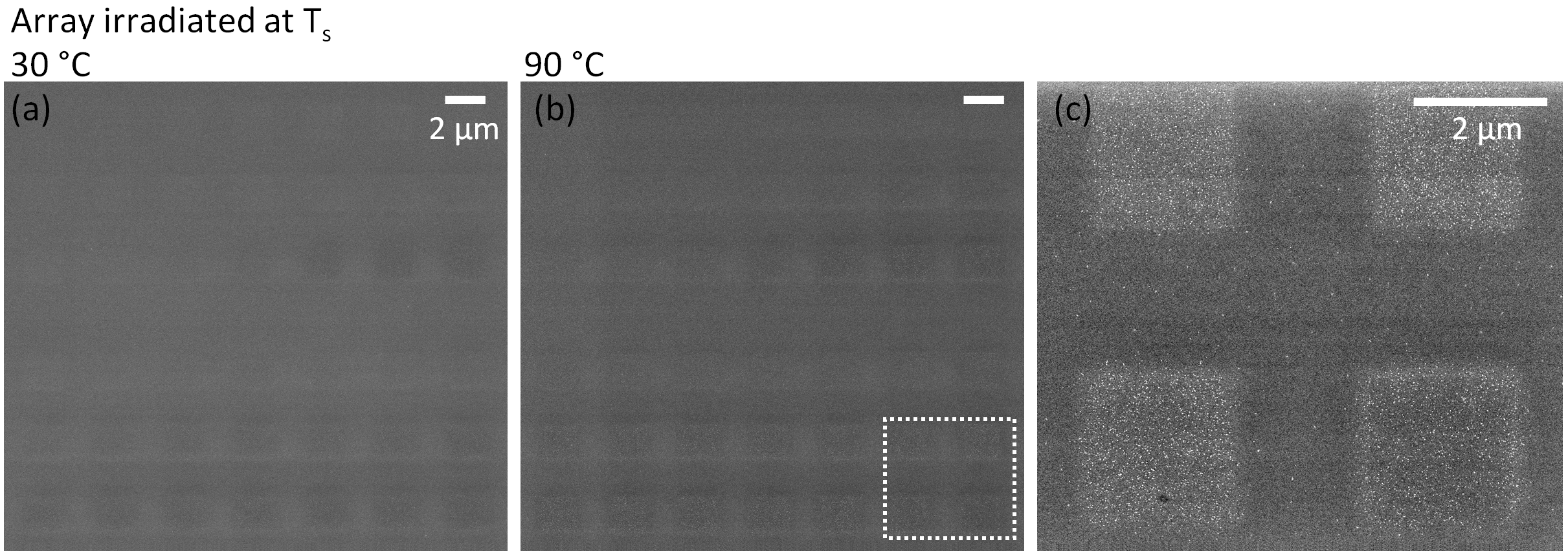}
\caption{SE micrographs of the arrays were captured after the control experiment. The arrays were irradiated with the electron beam at substrate temperatures of 30~\textdegree C (a) and 90~\textdegree C (b) without gold precursor flow. Following irradiation, the arrays underwent exposure to the gold precursor accompanied by heating to 90~\textdegree C for a duration of 50 min. A close-up high resolution SE micrograph (c) of the highlighted area in (b) reveals a small quantity of autocatalytically grown gold crystals within the irradiated regions.}
\label{SIfig:control1st}
\end{figure}

The first control test of pattern array deposition was performed without the Au(acac)Me$_2$ precursor gas being introduced into the vacuum chamber.
The array pattern was irradiated with the electron beam while residual chamber gas was present at a base pressure of $7\cdot10^{-7}$ mbar, both at room temperature (30~\textdegree C) and at 90~\textdegree C. 
Subsequently, both irradiated areas were left at elevated temperature of 90~\textdegree C for 50~min with the gold precursor gas flow enabled. 
SE micrographs presenting the results are displayed in Figure~\ref{SIfig:control1st}(a) and (b).
Gold particles are observed only for the highest dose as evidenced by the close-up high resolution SE micrograph provided in Figure~\ref{SIfig:full1st}(c), which is a magnified view of the highlighted region in Figure~\ref{SIfig:full1st}(b).
This implies that under high-vacuum (HV) conditions, surface activation by electrons, as also demonstrated by Walz et al.~\cite{Walz2010}, may be achievable.
Dark contrast in squares may arise from hydrocarbon contamination on the surface caused by residual gas composition in high vacuum~\cite{Rykaczewski2007}.
These deposits are commonly referred to as carbonaceous contamination. 

\begin{figure}[ht]
\centering
\includegraphics[width=15cm]{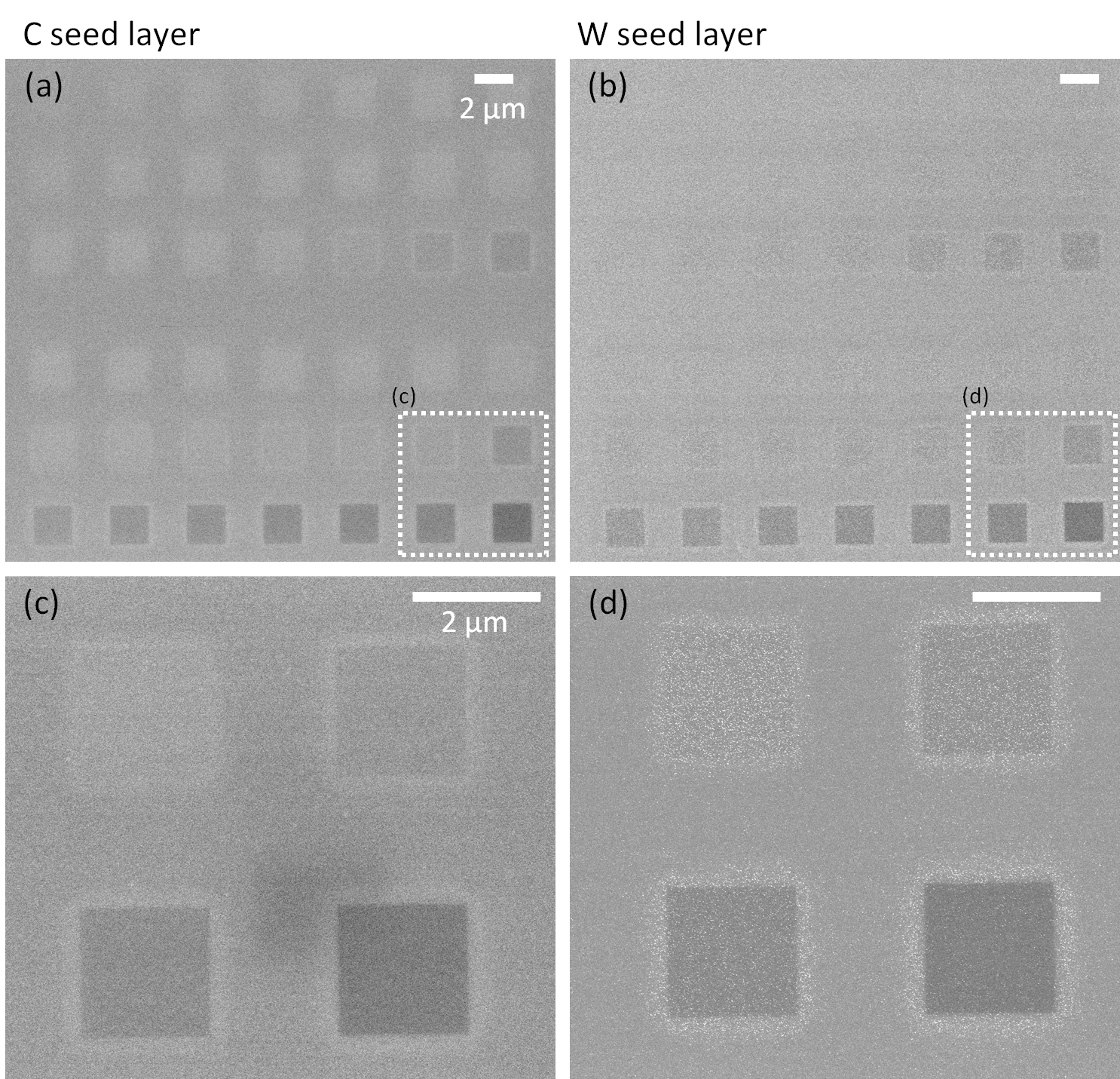}
\caption{\textcolor{black}{SE micrographs of the C (a) and W (b) seed layers, deposited at 90 \textdegree C captured after the control experiment. Following seeding, the arrays underwent exposure to the gold precursor accompanied by heating to 90~\textdegree C for a duration of 30 min. A close-up high resolution SE micrograph (c) and (d) respectively of the highlighted areas in (a) and (b) reveals negligible quantity of autocatalytically grown gold crystals within the irradiated regions.}}
\label{SIfig:control1st_cnw}
\end{figure}

\newpage
\textcolor{black}{The second control experiment was done with the seeding by carbon (C$_{10}$H$_{8}$ - naphthalene) and tungsten (W(CO)$_6$ - tungsten hexacarbonyl) precursors at 90~\textdegree C substrate temperature. 
Deposited arrays were heated to 90~\textdegree C, while being exposed to the gold precursor for 30 min.
The results are shown in Fig. \ref{SIfig:control1st_cnw}, and depict no significant amount of gold grown after the experiment in both cases.}

\newpage

\begin{figure}[ht]
\centering
\includegraphics[width=15cm]{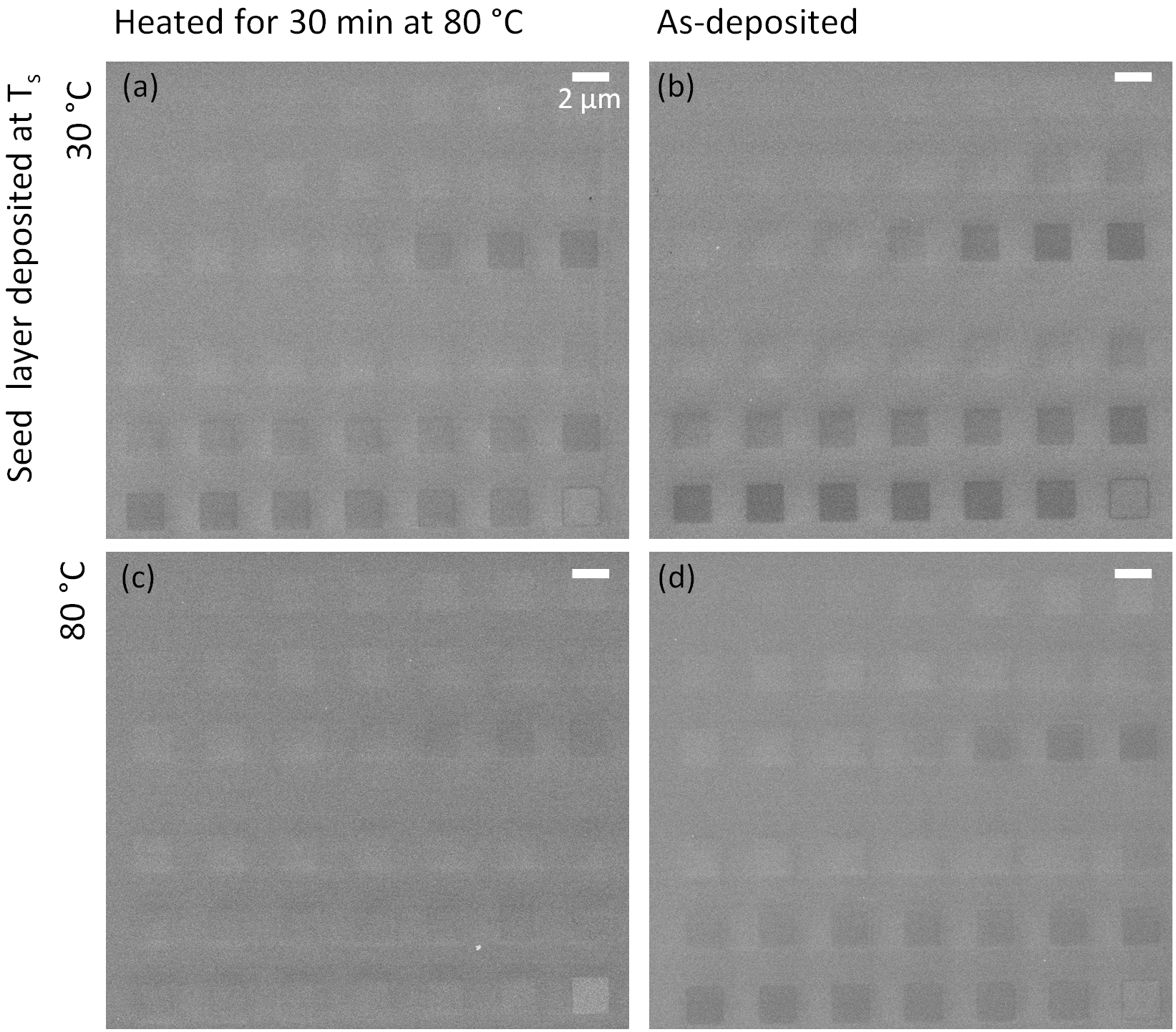}
\caption{\textcolor{black}{SE micrographs of the seed layers, deposited at 30~ \textdegree C (top row) and 80~\textdegree C (bottom row). Following seeding, two arrays were heated to 80~\textdegree C for a duration of 30 min without precursor flooding, and the results are represented in sub-figures (a) and (c). Sub-figures (b) and (d) show as-deposited seed arrays at corresponding substrate temperatures 30~\textdegree C and 80~\textdegree C for the comparison.}}
\label{SIfig:control1st_noprecursor}
\end{figure}

\textcolor{black}{The last control experiment was done with seeding by Au(acac)Me$_2$ precursor at 30 and 80~\textdegree C substrate temperature.
Test heating without precursor gas flow was then done at 80~\textdegree C for 30 min and then directly compared to as-deposited seed layers at corresponding temperatures. 
The results are shown in Fig. \ref{SIfig:control1st_noprecursor}, and depict no gold growth on the seeds at both temperatures in comparison to the as-deposited seed layers.}

\newpage
\section{Thickness Evaluation}

\begin{figure}[ht]
\centering
\includegraphics[width=17cm]{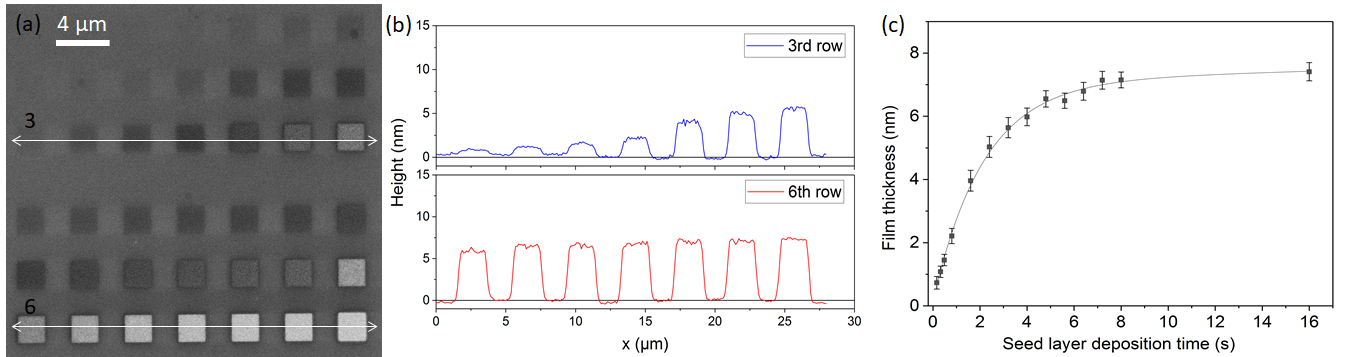}
\caption{AFM analysis was conducted on the gold films, which were seeded and underwent autocatalytic growth for 30 minutes at 80~\textdegree C. The resulting gold films were then examined using (a) SEM, and (b) AFM profiles of the 3rd and 6th rows were obtained. (c) The film thickness of each individual square in the 3rd and 6th rows was measured against seed layer deposition time.}
\label{SIfig:AFM}
\end{figure}

Atomic force microscopy (AFM) measurements were carried out to determine the thickness of the autocatalytically grown gold films. 
An SE image of the autocatalytically grown films at an elevated temperature of 80~\textdegree C for 30 min is shown in Figure~\ref{SIfig:AFM}(a). 
Seeding for the films was done at 80~\textdegree C. 
Figure~\ref{SIfig:AFM}(b) shows the AFM thickness profiles of the squares in the 3rd and 6th rows, which were seeded with a minimum pitch of 5 nm. 
The thickness of the film versus the deposition time of the seed layer for each individual square is displayed in Figure~\ref{SIfig:AFM}(c).
The data was compared to an exponential asymptotic function that exhibits a saturation of thickness.
The saturation in autocatalytic CVD gold growth was observed on certain seed squares with a deposition time of 5.6~s for a size of 2~\textmu m x 2~\textmu m, corresponding to an electron dose of 140~pC/\textmu m$^2$ or about 875 electrons per nm$^2$.
Corresponding thickness of such a film is about 6.5~nm.
Gold films after seeding and autocatalytic growth at 80~\textdegree C for longer times of 60 and 90 mins were additionally investigated. 
A saturation in achievable film thickness in both cases was also observed as discussed in the main manuscript.
Autocatalytic growth for 60~min only resulted in additional 0.7~nm, and even 90 minutes total growth time resulted only in 8.5~nm.

\begin{figure}[ht]
\centering
\includegraphics[width=15cm]{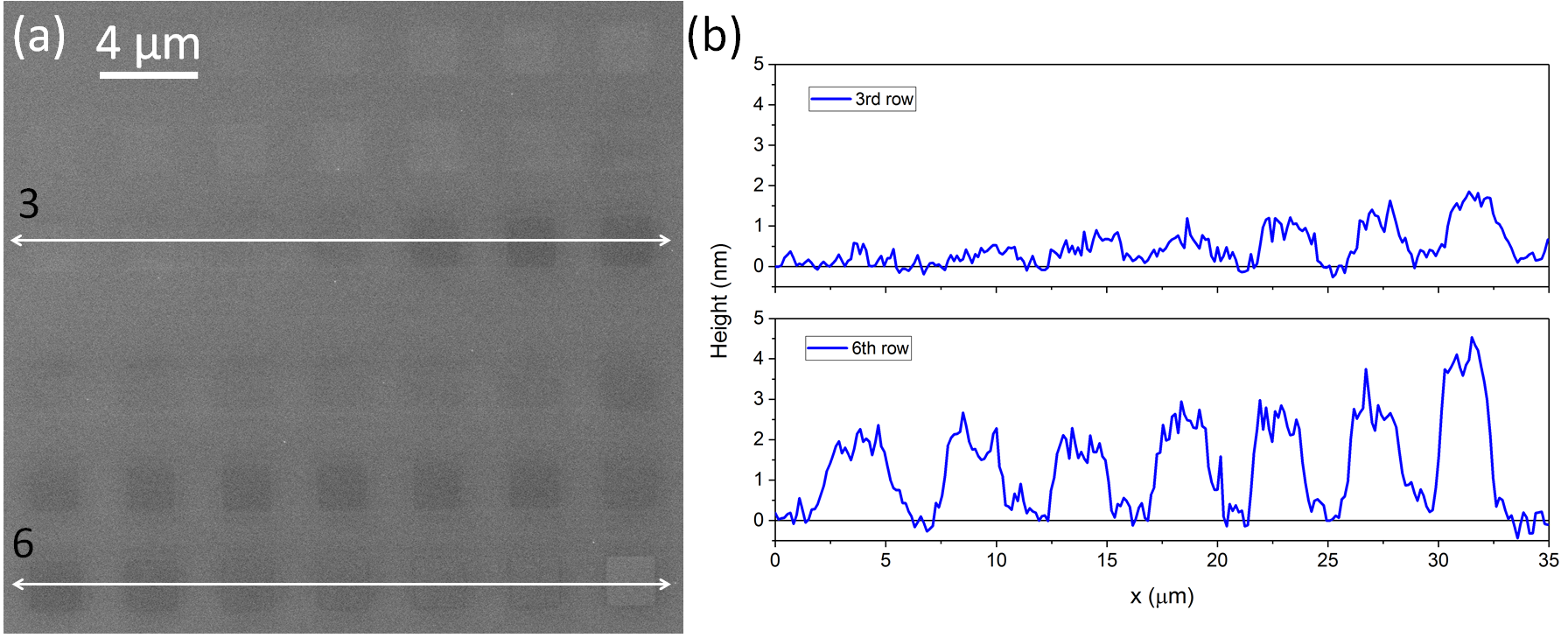}
\caption{\textcolor{black}{(a) SE micrograph of a seed layer, as-deposited onto 90~\textdegree C substrate. (b) AFM of the as-deposited seed layer.}}
\label{SIfig:afm_seed}
\end{figure}

\textcolor{black}{Additionally, a gold seed layer was investigated in AFM, which was deposited at 90~\textdegree C substrate temperature. 
SE micrographs as well as AFM profiles are depicted in Fig.~\ref{SIfig:afm_seed}.}

\newpage
\section{Transmission Electron Microscopy}

\begin{figure}[h!]
\centering
\includegraphics[width=19cm]{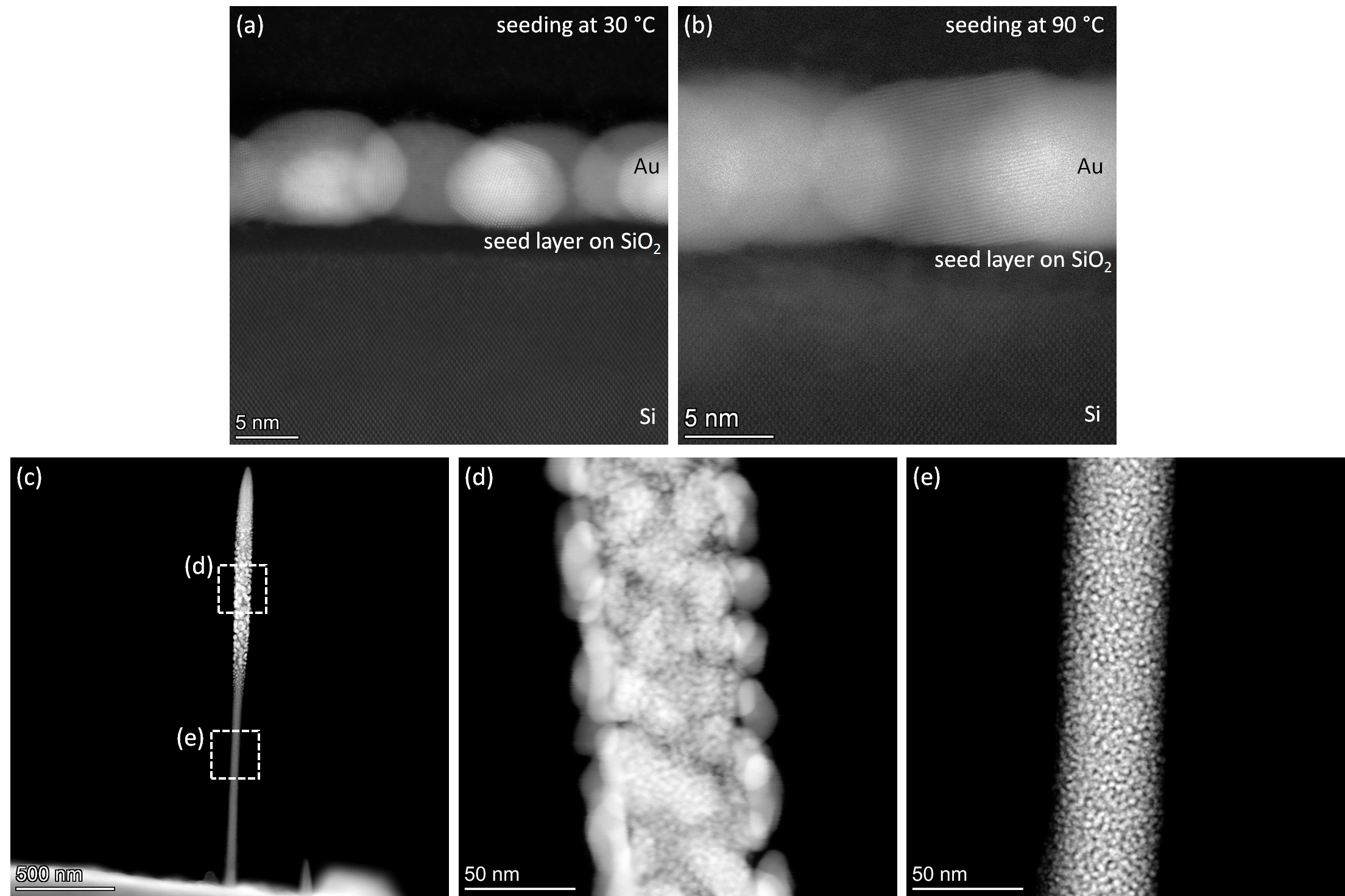}
\caption{STEM HAADF images of gold films grown on different silicon substrates with a native oxide layers at varying thicknesses are presented. The films were grown at 90~\textdegree C and with a seed layer deposited on a cold Si substrate of n-type at 30~\textdegree C (a) and at 90~\textdegree C of p-type Si chip (b). \textcolor{black}{STEM HAADF image of a nanopillar (c), grown by the electron beam induced deposition with autocatalytically grown gold crystals on its surface in the top part (d). Close up of STEM HAADF image of FEBID grown nanowire from the bottom part of the pillar (e).}}
\label{SIfig:TEM_full}
\end{figure}

For transmission electron microscopy (TEM) investigations, two different samples were used.
Gold films were grown autocatalytically on substrates at a temperature of 90~\textdegree C for 30 min, but were seeded at 30~\textdegree C for sample (a) and 90~\textdegree C for sample (b) (Figure~\ref{SIfig:TEM_full}(a) and (b) respectively). 
Gold films were grown on a n-type (100) Si chip with native oxide for sample (a) and on a p-type (100) Si chip with native oxide for sample (b).
TEM lamellae were prepared according to the standard lift-out protocol using dual beam FIB-SEM Tescan Lyra3.
The TEM images were captured using a probe-corrected ThermoFisher Scientific Titan Themis 200 G3. 
Figure~\ref{SIfig:TEM_full} shows the full STEM micrographs used for the main manuscript.
In both instances, periodic atomic order of specific gold grains can be observed.

\textcolor{black}{Additionally, nanopillars were grown on a standard FIB-TEM Cu grid to investigate crystallinity of gold grains, described in Section 3 of the main manuscript. 
Fig.~\ref{SIfig:TEM_full}(c) contains a HAADF TEM image of one pillar together with a close-up images from highlighted area (d) with gold crystals (grown autocatalytically) and (e) FEBID grown pillar.
Due to the mechanical vibrations, it is impossible to resolve periodic atomic pattern within crystals.
}
\newpage

\section{Monte-Carlo Simulations}

\begin{figure}[ht]
\centering
\includegraphics[width=16cm]{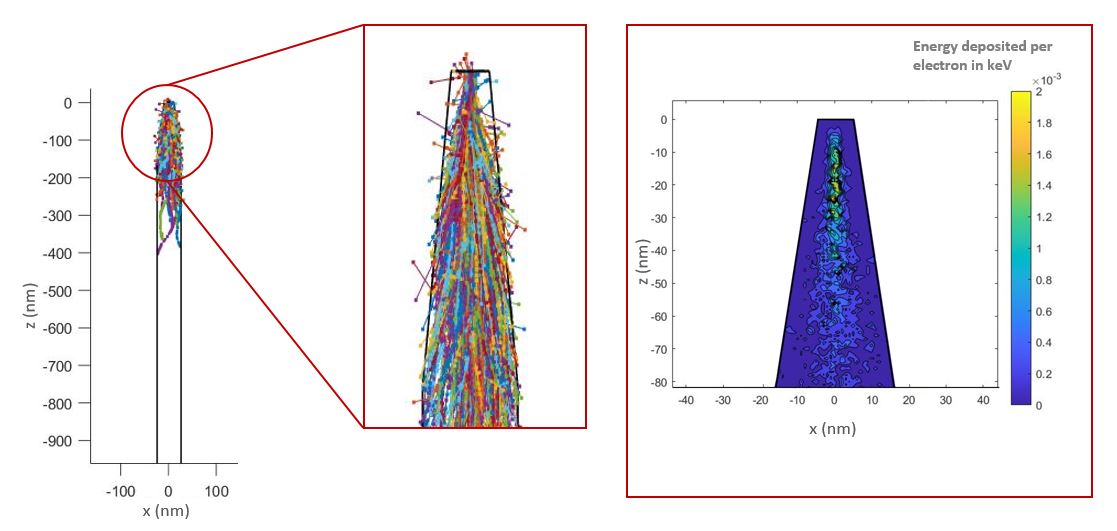}
\caption{The trajectory of electrons with a primary energy of 15~keV inside the column with 10~nm upper radius of the truncated cone with 150~nm height and 25~nm cylinder. The inset shows the trajectories near the tip and the contour plot of the energy deposited per electron and line segment between two collisions.}
\label{SIfig:electron_trajectories}
\end{figure}

The energy deposited by electrons in the material was extracted using a Monte-Carlo method proposed by David C. Joy's single scattering model~\cite{Joy1995}.
The analytical Mott expression~\cite{Browning1994} implemented the inelastic scattering event, and the modified Bethe expression~\cite{Joy1995} was used for the elastic energy loss, which is an average energy loss by an electron due to all inelastic scattering events.
The structure representing the FEBID-grown pillar was implemented with a cylinder and a truncated cone above the cylinder. 
Trajectory and energy loss of each electron were recorded and used to calculate the total energy lost by primary electrons on average. 
Figure~\ref{SIfig:electron_trajectories} depicts the paths of 500 electrons with 15~keV energy, traveling within an object that has a top truncated cone with a 10~nm radius and a height of 150~nm, as well as a cylinder with a 25~nm radius. 
The object has a composition ratio of Au:C as 0.28:0.72, with a density of 6.5~g/cm$^3$. 
The energy transfer value for 4 g/cm$^3$ lower bound was calculated using the same method.
The object's geometry was derived from SEM images.
The energy deposited per each primary electron and line segment between two collisions at the apex of the pillar is shown in the inset of the figure.

\section{Temperature Distribution in Pillars Induced by the Electron Beam}

Convection losses in high vacuum ($6\cdot10^{-7}$~mbar) are negligible because the mean free path of the residual gas molecules is greater than the size of the vacuum chamber.
The radiative losses can be estimated by applying the Stefan-Boltzmann law, where the total radiative power $P_r$ is the difference between the power emitted by the column and the power absorbed by the surrounding medium: $P_r~=~A\sigma\varepsilon\big(T^4-T_0^4\big)$, where $A$ is the surface area of the column, $\sigma~\approx~5. 67\cdot10^{-8}$~W/m$^2$K$^4$ is the Stefan-Boltzmann constant, $\varepsilon$ is the emissivity, and $T_0$ is the ambient temperature of the vacuum chamber. 
The typical pillar deposited at 30~\textdegree C has a surface area of $A~=~2\pi r_p l_p~\approx~4.24\cdot10^{-13}$~m$^2$ with a radius of $r_p~=~25$~nm and a length of $l_p~\approx~2700$~nm. 
In order to estimate the maximum radiative losses, we assume a maximum emissivity of the deposit of 1.
For the aforementioned reason, we ignore any irregularities in temperature distribution and assume uniform heating of the entire pillar, with its length being equal to the previously determined temperature of 112~\textdegree C (385~K).
In this instance, the maximum power output is $P_r~\leq~3.87\cdot10^{-10}$~W. 
As per Monte-Carlo simulations, a primary electron of 15~kV conveys at least 0.75~kV of its initial energy to the pillar.
This then results in an implanted power of $P_i~=~\Delta E_{Joule}I~= 7.5\cdot10^{-8}$~W by a focused electron beam with a current of 100~pA.
If the power transfer from the focused electron beam is compared with the maximum estimated radiative loss, the latter is at least 2 orders of magnitude lower and thus negligible, which justifies a linear temperature gradient in the pillar.

\begin{figure}[ht]
\centering
\includegraphics[width=16cm]{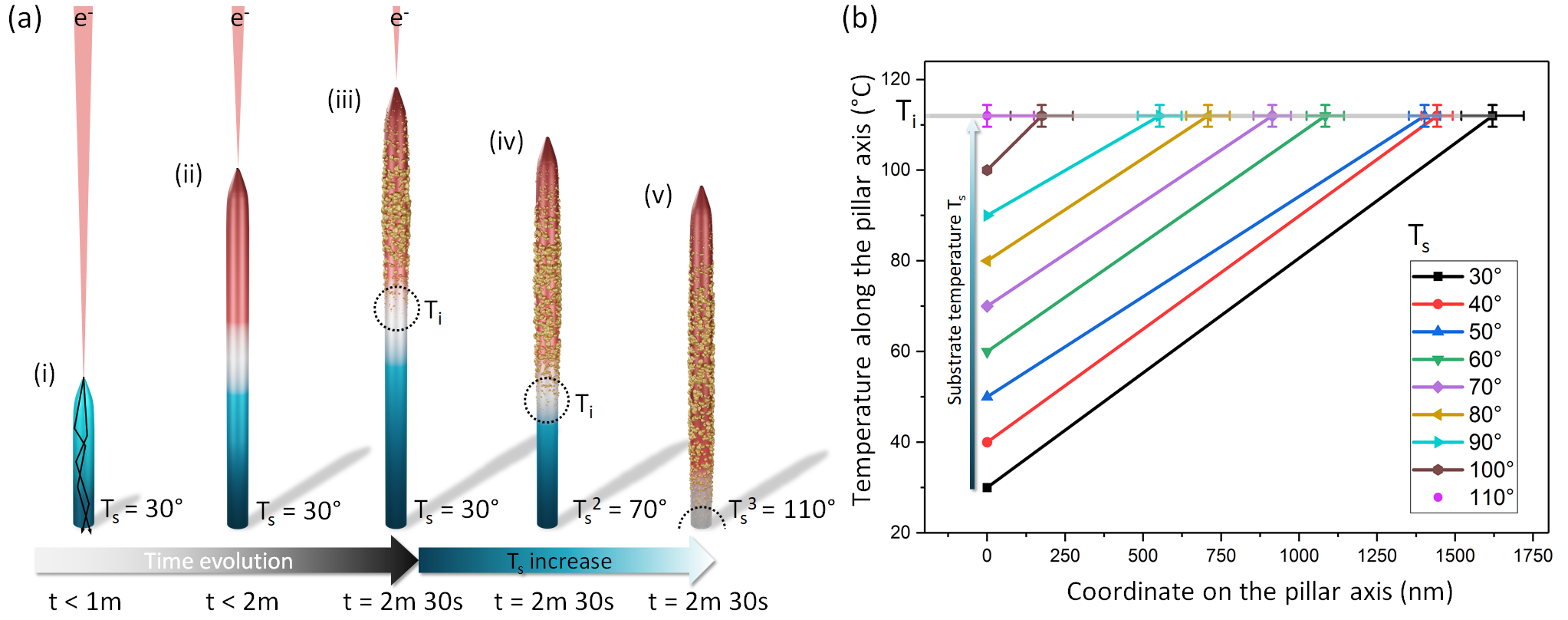}
\caption{Pillar growth schematics (a). The time evolution shown on the left shows the pillar during FEBID growth below the 1~min mark (i), where heat transfer to the pillar is not maximised. The black lines indicate electron trajectories within the pillar. The forward scattered trajectories that are outside the pillar are not shown, as they have no contribution to the temperature inside the pillar. Later, heat transfer becomes saturated, and a linear temperature distribution sets in (ii). However, it takes a minimum of 2 min 30 s for autocatalysis to initiate. Autocatalytic gold particles can only become observable on the top section of the pillar at that time (iii). With a constant deposition time and an increase in sample temperature T$_s$, the point where autocatalysis starts on the pillar moves towards the sample surface (iii)-(v). Linear temperature distribution in the pillars (b) was observed experimentally when they were deposited on the substrate at various elevated temperatures ranging from 30~\textdegree C to 110~\textdegree C.} 
\label{SIfig:pillars}
\end{figure}

The following schematics are proposed for a better understanding of the heat transfer and temperature distribution in the electron beam grown pillars. 
The left-hand side of Figure~\ref{SIfig:pillars}(a) depicts the time evolution of a single pillar.

Let us first consider the stage temperature T$_s$ set at the room temperature of
30~\textdegree C. At smaller sizes, the majority of the heat supplied via electron beam is dissipated within the substrate. 
However, once the pillar reaches a certain height that corresponds to the stopping range of primary electrons in the rod, the implanted energy from primary electrons into the pillar will be maximised.
The temperature of the top of the pillar exceeding the base temperature (i.e. the temperature of the substrate) generates an almost linear temperature gradient along the axis of the pillar.
This can be attributed to a quasi-continuous heat supply by the electron beam and the stage temperature, T$_s$, remaining constant, resulting in a steady solution of the heat transfer equation (cf. \cite{Reimer2000}).
This occurs because there are no significant convectional or radiative losses as estimated previously.
As a result, the temperature of the upper portion of the pillar may reach the threshold for the autocatalytic growth of gold particles, which only becomes visible once the gold crystals increase to a size that can be observed by SEM imaging at the sidewall of the pillar.
During the experiment with Au(acac)Me$_2$, this occurred at about 2 min 30 s with a 15~keV 100~pA electron beam.
The temperature distribution within the upper section of the pillar remains uncertain because of the presence of gold crystals, which impact heat conductivity.

If a pillar is now deposited with the same deposition parameters and total time, but on the stage at an elevated temperature of 70~\textdegree C, an autocatalytic process will take place at a lower point along the pillar axis (see Figure~\ref{SIfig:pillars}(a)-(iv)).
By progressively increasing the stage temperature, gold crystals will eventually appear on the pillar alongside the sample surface, as shown in Figure~\ref{SIfig:pillars}(a)-(v).
This enables the calculation of T$_i$ - the temperature induced by the electron beam at the point along the pillar where the CVD process first occurs (see Figure~\ref{SIfig:pillars}(a)).
T$_i$ remains constant during the specified deposition period.
If we plot all temperature distributions for each pillar grown at different T$_s$ (refer to Figure~\ref{SIfig:pillars}(b)), we can observe that the point on the pillar heated to the temperature T$_i$ shifts linearly to the sample surface and collapses with the surface at around 112~\textdegree C. 

\newpage
\section{\textcolor{black}{Line width experiment}}

\begin{figure}[ht]
\centering
\includegraphics[width=12cm]{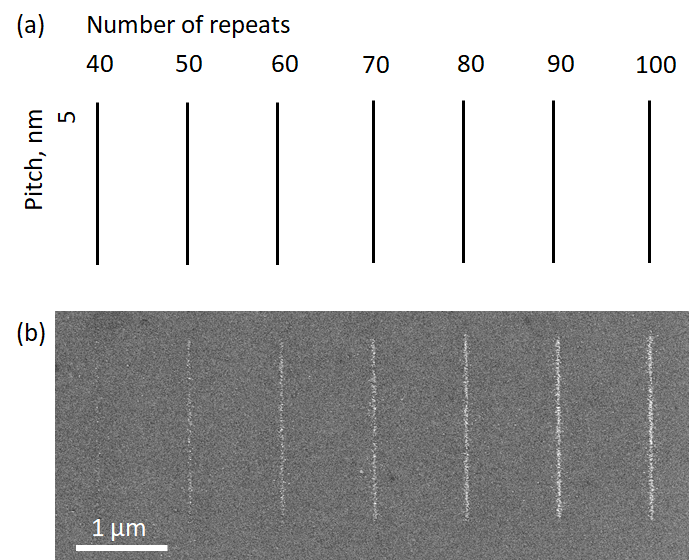}
\caption{\textcolor{black}{SE micrograph of gold lines (b) grown autocatalytically by CVD for 30 min on the respective seed lines (a). Both seeding and subsequent autocatalytic gold CVD growth were done at 90~\textdegree C substrate temperature.}}
\label{SIfig:linewidth}
\end{figure}

\textcolor{black}{An additional line width experiment was conducted. The seed layer with a simple line pattern (as shown in Fig. \ref{SIfig:linewidth} (a)) was deposited onto a 90~\textdegree C substrate, followed by a standard autocatalytic growth of gold at 90~\textdegree C. The results are presented in Fig. \ref{SIfig:linewidth} (b). For 5~nm pitch patterning with 5~kV and 100~pA electron beam, a saturation in growth occurs at 80 repeats and reaches approximately 40~nm.}

\textcolor{black}{The actual electron beam probe size is much smaller (below 5~nm) than the obtained line width. 
The lateral structure widening is caused by SE and BSE and strongly depends on the acceleration voltage of the beam as known from e-beam lithography literature. 
The best area-selective fidelity could be achieved by setting the lowest available primary beam voltage to lower than 1~kV. 
Experimentally, at 5~kV we observed the broadening to be less than 100 nm, which can be seen in the inset (c) and (d) of Fig. 2.(b) of the main manuscript. 
}

\medskip


\bibliographystyle{MSP}
\bibliography{MSP-template}

\end{document}